\documentclass[12pt,letterpaper]{article}
\usepackage[a4paper, total={7in, 10in}]{geometry}

\usepackage{graphicx}
\usepackage{helvet}
\usepackage{authblk}
\usepackage{hyperref}
\usepackage{amsmath} 
\usepackage{amssymb} 
\usepackage{orcidlink} 
\usepackage[super,comma,sort&compress] 
    {natbib}
\usepackage[right,mathlines]{lineno}
\usepackage[title]{appendix}%
\usepackage[utf8]{inputenc} 
\usepackage[T1]{fontenc}    
\usepackage{url}  
\usepackage{booktabs}  
\usepackage{amsfonts}  
\usepackage{nicefrac}  
\usepackage{adjustbox}
\usepackage{longtable}
\usepackage{microtype} 
\usepackage{xcolor}    
\usepackage{multirow}
\usepackage{threeparttable}
\usepackage{amsmath}
\usepackage{subcaption}
\usepackage{makecell}
\usepackage{placeins}
\usepackage[utf8]{inputenc} 
\usepackage{booktabs} 
\usepackage{siunitx}
\usepackage{caption}
\usepackage{tabularx}
\usepackage{listings}   
\usepackage{pifont}
\usepackage{array}
\usepackage{makecell}
\usepackage{colortbl}
\usepackage{xcolor}
\usepackage{siunitx} 
\usepackage[capitalize]{cleveref}
\theadset{\bfseries}

\makeatletter
\renewcommand{\maketitle}{\bgroup\setlength{\parindent}{0pt}
\begin{flushleft}
  \textbf{\@title}
  
  \@author
\end{flushleft}\egroup}
\makeatother

\title{AnyECG: Evolved ECG Foundation Model for Holistic Health Profiling}
\date{}

\author[1,2,\#\orcidlink{}]{Jun Li}
\author[5,\#]{Hongling Zhu}
\author[1,2,\#]{Yujie Xiao}
\author[6,\#]{Qinghao Zhao}
\author[7,8,9]{Yalei Ke}
\author[1,2]{Gongzheng Tang}
\author[1,2,10]{Guangkun Nie}
\author[11]{Deyun Zhang}
\author[12]{Jin Li}
\author[7,8,9]{Canqing Yu}
\author[1,2,3,4,*,\orcidlink{0000-0001-7521-5127}]{Shenda Hong}

\affil[1]{National Institute of Health Data Science, Peking University, Beijing, China}
\affil[2]{Institute of Medical Technology, Health Science Center of Peking University, Beijing, China}
\affil[3]{Institute for Artificial Intelligence, Peking University, Beijing, China}
\affil[4]{NHC Key Laboratory of Cardiovascular Molecular Biology and Regulatory Peptides, Peking University, Beijing, China}
\affil[5]{Division of Cardiology, Department of Internal Medicine, Tongji Hospital, Tongji Medical College, Huazhong University of Science and Technology, Wuhan, Hubei, China}
\affil[6]{Department of Cardiology, Peking University People’s Hospital, Beijing, China}
\affil[7]{Department of Epidemiology and Biostatistics, School of Public Health, Peking University, Beijing, China}
\affil[8]{Peking University Center for Public Health and Epidemic Preparedness \& Response, Beijing, China}
\affil[9]{Key Laboratory of Epidemiology of Major Diseases (Peking University), Ministry of Education, Beijing, China}
\affil[10]{School of Intelligence Science and Technology, Peking University, Beijing, China}
\affil[11]{HeartVoice Medical Technology, Hefei, China}
\affil[12]{Division of Cardiology, Department of Internal Medicine, Tongji Hospital, Tongji Medical College, Huazhong University of Science and Technology, Wuhan, Hubei, China}

\affil[$\#$]{These authors contributed equally}
\affil[*]{Correspondence: hongshenda@pku.edu.cn}

\begin{document}

\maketitle

\section*{ABSTRACT}

Background: Artificial intelligence-based electrocardiography (AI-ECG) has been proven to detect a variety of cardiac and non-cardiac pathological changes. However, current mainstream research remains limited to independent models for identifying single diseases. These models insufficiently address complex clinical comorbidities and rarely incorporate predictions of future disease risk. Although our previous work, ECGFounder, has achieved coverage of 150 cardiac diseases and significantly improved model performance through fine-tuning techniques, further breakthroughs and expansions are still needed to address key clinical needs and achieve holistic health profiling such as more comprehensive disease screening, long-term risk prediction, and in-depth comorbidity pattern recognition.

Methods: In this study, we constructed a large-scale, multicenter electrocardiogram (ECG) dataset containing 13,348,593 ECG records and their corresponding ICD diagnostic codes from 2,984,209 patients. This dataset integrates retrospective and prospective data, providing a data foundation for multiple tasks including current disease diagnosis, future risk prediction, and comorbidity identification. Methodologically, we employed a transfer learning strategy to fine-tune the pre-trained ECGFounder, successfully developing an improved ECG foundation model, AnyECG, aimed at significantly enhancing its capabilities in holistic health profiling. Furthermore, we conducted extensive multicenter external validation and comprehensively evaluated the model's robustness and accuracy in current diagnosis, future prediction, and comorbidity identification in a 10-year longitudinal cohort study.

Results: We explored the systemic prediction ability of AnyECG across 1,172 ICD-coded conditions. AnyECG achieved an AUROC above 0.7 for 306 of these diseases. The AnyECG model reveals many novel discoveries that have never been reported, demonstrates robust detection of comorbidity patterns, and shows predictive capability for diseases that may emerge in the future. For example, AnyECG achieved surprising diagnostic abilities for hyperparathyroidism (AUROC: 0.941 [95\% CI: 0.903–0.977]), type 2 diabetes mellitus (AUROC: 0.803 [95\% CI: 0.798–0.807]), Crohn’s disease (AUROC: 0.817 [95\% CI: 0.774–0.857]), lymphoid leukemia (AUROC: 0.856 [95\% CI: 0.849–0.861]) and chronic obstructive pulmonary disease (AUROC: 0.773 [95\% CI: 0.759–0.786]). 

Conclusion: This study introduces AnyECG, a new foundation model, and provides substantial empirical evidence for the clinical application of AI-ECG in both current disease detection and future disease prediction, highlighting its potential as a systemic diagnostic tool.


\section*{KEYWORDS}

ECG, Foundation Model, Disease Profiling

\section*{INTRODUCTION}
Electrocardiogram (ECG) is a non-invasive, cost-effective, and widely available tool traditionally used to assess cardiac electrical activity \cite{stracina2022golden,liu2021deep}. Abnormalities in electrical cardiac activity affect millions of people worldwide and are increasingly recognized as indicators and contributing factors for various health conditions \cite{lippi2021global}. It is worth noting that subtle changes in ECG signals often precede the clinical onset of diseases, including arrhythmias, heart failure, and ischemic heart disease \cite{wang2023early,kim2025electrocardiography,attia2023explainable}. These associations highlight the central role of cardiac electrophysiological homeostasis in maintaining overall health and underscore the potential of ECG for early prediction of a wide range of diseases \cite{turgut2025unlocking,liu2021deep}. However, most existing studies rely solely on isolated ECG indicators (such as QT interval or ST-segment deviations) to establish their associations with specific diseases \cite{deng2020association,buszman1995use}, while the complex physiological information embedded in ECG signals has yet to be fully explored and integrated for analysis.\cite{chen2022review} 

In recent years, the integration of artificial intelligence (AI) with ECG—termed AI-enhanced ECG (AI-ECG)—has demonstrated significant potential in diagnosing and predicting various diseases beyond arrhythmias, including heart failure, chronic kidney disease, and diabetes \cite{attia2021external,dhingra2025artificial,holmstrom2023deep,kim2024deep}. While these efforts represent substantial progress, they face several critical limitations \cite{chung2022clinical,siontis2021artificial}. Most methods target specific diseases or limited disease categories and are constrained by relatively small datasets \cite{moreno2024ecg}. Furthermore, existing models lack large-scale multi-center external validation, limiting their generalizability to real-world clinical settings \cite{moosavi2024prospective,karabayir2025generalizability}. Additionally, most studies focus solely on diagnosis and do not conduct longitudinal cohort-based prediction and risk assessment. Moreover, prior research has not systematically explored the scientific discoveries underlying ECG's ability to diagnose and predict diseases \cite{muzammil2024artificial}.

Foundation models offer a promising solution to address these challenges through transfer learning. In the biomedical field, foundation models have exhibited remarkable capabilities across multiple medical domains, advancing disease prediction, diagnosis, and therapeutic discovery \cite{moor2023foundation,liu2025biomedical}.
Recently, several ECG foundation models have been proposed \cite{mckeen2024ecg,song2024foundation,li2025electrocardiogram}. ECG-FM was pre-trained using self-supervised learning on two large ECG datasets \cite{mckeen2024ecg}. CREMA was designed to learn generalizable representations through self-supervised pretraining on 1 million ECGs dataset.\cite{song2024foundation}
Despite achieving commendable results, these studies on ECG foundation models narrowly focus on arrhythmia- and cardiac-specific outcomes, restricting the broader potential of foundation models in disease prediction.
In our earlier research, we developed ECGFounder using a supervised learning paradigm, based on the largest current ECG dataset—the Harvard-Emory ECG Database (HEEDB) \cite{li2025electrocardiogram,koscova2024harvard}. This dataset included 10,771,552 ECGs from 1,818,247 patients. While ECGFounder performed well on several downstream tasks, establishing the feasibility of our AI-ECG foundational model, its core capabilities remain primarily focused on arrhythmia detection. Further research is needed to explore how to leverage ECG for screening non-cardiac systemic diseases, predicting future health risks, and identifying comorbidity patterns—i.e., constructing holistic health profiling.

In this study, we systematically explored the role of cardiac electrical activity in disease diagnosis and risk prediction both in cardiac or non-cardiac diseases. We fine-tuned a pre-trained ECGFounder model based on large-scale paired ECG–ICD data to enhance its ability to identify multiple disease spectrums, and performed external validation in multi-center independent cohorts. To further examine the model's cross-population and cross-regional generalization performance, we also conducted additional validation in the China Kadoorie Biobank (CKB) \cite{chen2011china}. Overall, this study constructed a large-scale, multicenter, multinational ECG cohort covering 2,984,209 patients and 13,348,593 ECGs, including both retrospective and prospective data, providing a data foundation for creating holistic health profiles of patients.



This approach enables the model to align with various diseases and clinical events, ultimately developing AnyECG, a versatile model designed to predict over 800 disease types. Through extensive multi-center validation, we aim to demonstrate that AnyECG can (1) diagnose a wide spectrum of diseases, (2) analyze comorbidity relationships, and (3) perform effective disease screening (Figure 1b). Furthermore, we explored the correlations among different diseases within the ECG framework. By identifying association patterns across various disease systems, we hope to contribute to the understanding of disease mechanisms and potentially uncover new insights into disease interactions. 

Our objective was to further enhance ECGFounder using paired ECG and ICD diagnosis data, and to develop a unified model, AnyECG, covering the full spectrum of 1,172 diseases. With this model, we reported previously unrecognized disease associations detectable from ECGs, demonstrated its ability to reveal links between comorbidities, and showed its potential for predicting future disease onset. In this way, we extend the role of the ECG from a cardiac-centered tool to a comprehensive platform for current disease detection and future disease prediction.

\begin{figure*}
    \centering
    \includegraphics[width=\textwidth]{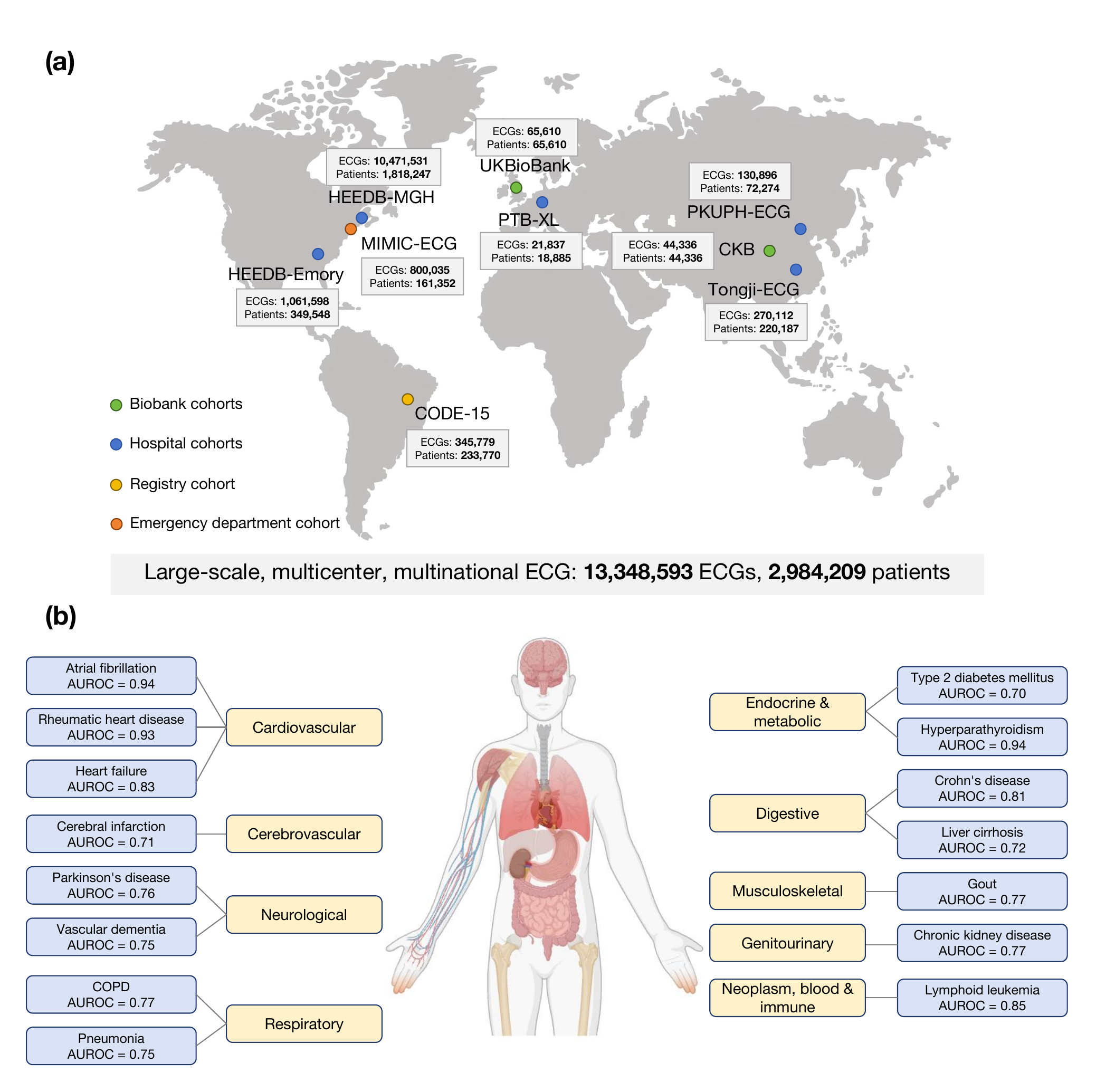} 
    \caption{\textbf{Overview of the development and results of AnyECG.} 
(a) The source and composition of the dataset. The dataset we used included large-scale, multicenter, multinational ECG cohorts: 13,348,593 ECGs, 2,984,209 patients. 
(b) Representative results of AnyECG across different systems are presented. Outcome indicators for different diseases are shown across nine systems. }
    \label{fig:combinedframework}
    \vspace{2mm}
    \hrule
\end{figure*}

\section*{Methods}

\subsection*{Datasets}

We utilized nine large-scale ECG Datasets, MGH-ECG Dataset \cite{koscova2024harvard}, Emory-ECG Dataset \cite{koscova2024harvard}, PTB-XL \cite{ptbxl}, CODE-15 \cite{ribeiro2021code}, MIMIC-ECG \cite{gow2023mimic}, and Huazhong University of Science \& Technology Tongji Hospital Dataset (Tongji-ECG) for model development, and Peking University People's Hospital ECG Dataset (PKUPH-ECG), UKBB-ECG \cite{ukbb} together with the China Kadoorie Biobank ECG Dataset (CKB-ECG) for external validation \cite{chen2011china}.  

\textbf{HEEDB-MGH Dataset} The HEEDB-MGH Dataset is a large-scale, retrospective collection of 12-lead ECGs acquired in routine clinical care, developed through collaboration between Harvard University and Emory University \cite{koscova2024harvard}. It contains 10,608,417 ECG records from 1,818,247 patients at Massachusetts General Hospital (MGH), spanning several decades from the 1990s onwards. Records are sampled at either 250 Hz or 500 Hz, each with a duration of 10 seconds, and stored in WFDB format. Annotations generated by the 12SL-ECG analysis program are available for 10,471,531 MGH records, covering diagnostic statements, ECG morphology, and rhythm patterns. The dataset provides temporal and demographic diversity, making it a valuable resource for large-scale studies on cardiac diseases, arrhythmia prediction, and other ECG-related conditions.

\textbf{Tongji-ECG Dataset} The Tongji-ECG Dataset contains 270,112 ECGs with corresponding ICD-coded diagnostic information from 220,187 patients across three campuses of Tongji Hospital, Huazhong University of Science and Technology (Main Campus, Optical Valley Campus, and Sino-French New City Campus) in Wuhan, China. Data were collected between 2017 and 2024, with patient demographic characteristics summarized in Supplementary. All 12-lead ECGs were sampled at 500 Hz, with a duration of 10 seconds.

\textbf{PKUPH-ECG Dataset} The PKUPH-ECG Dataset comprises 130,896 ECGs from 72,274 patients, each with ICD-coded diagnostic information. Data span from 2013 to 2024, with patient demographics summarized in Supplementary. All ECGs were recorded using a 12-lead configuration, sampled at 500 Hz, with a duration of 10 seconds.

\textbf{CKB-ECG Dataset} The CKB-ECG Dataset is a large population-based prospective cohort of more than 500,000 participants aged 30–79 years, recruited from ten geographically diverse regions across China\cite{chen2011china}. Among them, 44,336 participants underwent 12-lead ECG recording as part of extensive phenotyping, with data linked to detailed demographic, lifestyle, environmental, and clinical information, as well as genomic data from genome-wide genotyping and whole-genome sequencing in selected subsets.

\textbf{MIMIC-ECG Dataset} The MIMIC-ECG (MIMIC-IV-ECG) Dataset is a large-scale collection of 800,035 diagnostic 12-lead ECG recordings from 161,352 patients treated at Beth Israel Deaconess Medical Center (BIDMC) in Boston, USA \cite{gow2023mimic}. These ECGs are linked to the MIMIC-IV electronic health record, providing rich clinical context including diagnoses, medications and laboratory measurements. The diagnostic ECGs were acquired as part of routine care across the emergency department, intensive care units and outpatient facilities, and cover the period from 2008 to 2019. All waveforms are standard 10-second, 12-lead recordings sampled at 500 Hz, enabling direct comparison with other contemporary clinical ECG datasets.

\textbf{PTB-XL Dataset} The PTB-XL Dataset is a large, freely accessible clinical ECG corpus curated by the Physikalisch-Technische Bundesanstalt (PTB) in Germany \cite{ptbxl}. It comprises 21,837 10-second 12-lead ECG recordings from 18,885 patients, with extensive cardiologist-verified diagnostic labels organized in a hierarchical taxonomy. Recordings were collected at PTB between October 1989 and June 1996 as part of long-term ECG research and later converted into a standardized Dataset. Each ECG is available at the original sampling frequency of 500 Hz (with an additional downsampled 100 Hz version), making PTB-XL a widely used benchmark for supervised and self-supervised ECG modeling.

\textbf{UKBB-ECG Dataset} The UKBB-ECG Dataset is derived from the UK Biobank, a large prospective cohort of around 500,000 participants recruited across the United Kingdom \cite{ukbb}. Digital 12-lead ECGs were acquired at-rest imaging visits, yielding ECG measurements in 65,610 individuals from the general UK population. UK Biobank Resting ECGs were recorded at dedicated UK Biobank Imaging Assessment Centres using GE Cardiosoft systems, stored as XML files (Data-Field 20205).Each record is a 10-second, 12-lead tracing sampled at 500 Hz with 5,000 data points per lead, providing high-resolution waveforms suitable for large-scale genetic and phenotypic association studies.

\textbf{CODE-15 ECG Dataset} The CODE-15 ECG Dataset is a large, population-based Dataset constructed as a stratified 15\% sample of the Brazilian CODE (Clinical Outcomes in Digital Electrocardiology) cohort \cite{ribeiro2021code}. It contains 345,779 12-lead ECG examinations from 233,770 patients, collected by the Telehealth Network of Minas Gerais across 800+ municipalities in the state of Minas Gerais, Brazil. ECGs were acquired during routine telehealth care between 2010 and 2016, with recordings stored as digital waveforms and linked to clinical reports. Signals are standard 12-lead recordings sampled at 400 Hz, with variable duration (a large subset of 143,328 exams providing 10-second segments), making CODE-15 one of the largest publicly available ECG datasets for developing and benchmarking deep-learning models for automatic ECG diagnosis and risk prediction.

\subsection*{Data Preprocessing}
The data preprocessing in this study followed a standardized and rigorous process. First, damaged, incomplete, or metadata-inconsistent records were removed. Then, all ECG signals with sampling frequencies below 500 Hz were resampled to 500 Hz using linear interpolation. Next, a 50/60 Hz notch filter was used to suppress power line interference, followed by a fourth-order Butterworth bandpass filter (0.67–40 Hz) to preserve physiologically relevant components while reducing high- and low-frequency noise. Finally, baseline drift was corrected by subtracting a 0.4-second median-filtered baseline estimate. This process ensured that the final signal retained the 0.67–40 Hz frequency band while minimizing noise, power line artifacts, and drift. For records longer than 10 seconds, consecutive 10-second segments were extracted; for shorter records, zero-padding was used. All signals were normalized using the mean and standard deviation of each segment before being input into the model.

\subsection*{Matching ECGs and ICDs}
To achieve systematic prediction of multiple disease spectrums based on ECG, we matched discharge ICD diagnoses in patients' electronic health records (EHR) with their corresponding ECG one-to-one, and constructed a scalable data integration workflow to form a high-quality ECG-ICD diagnosis matching queue.

At the coding level, we first standardized all ICD diagnoses, including removing leading zeros and standardizing the coding format, to minimize biases caused by differences in recording methods across centers and time periods. The cleaned ICD codes were further converted into multi-label binary matrices, allowing for a structured representation of all diagnostic information for each patient, thus adapting to the multi-label output requirements of deep learning models. To enhance clinical interpretability, we mapped all ICD codes to standardized diagnostic names and systematically cleaned up redundant labels caused by differences in coding rules.

In the data alignment stage, we synchronized ECG records with patient discharge times, ensuring correct alignment by matching time intervals and patient identifiers. Only patients with valid ECG data at discharge were included in the final analysis queue. The process ensures the temporal correlation and clinical consistency between diagnostic labels and ECG signals, thereby significantly improving the reliability and clinical significance of downstream model training.

Finally, we systematically grouped all paired data according to the first letter of the ICD code (i.e., the top-level diagnostic system), allowing the predictive performance of different disease systems to be evaluated independently. Through this series of rigorous coding standardization, label construction, data alignment, and hierarchical processing, we built a high-quality, multi-disease, multi-label large-scale ECG-diagnostic cohort, providing a solid data foundation for subsequent model training, generalization evaluation, risk prediction, and comorbidity identification.


\subsection*{AnyECG Develpment}
We construct an advanced deep learning architecture based on the Net1D framework and integrate an attention mechanism to enhance spatiotemporal feature extraction from ECG data \cite{hong2020holmes}. This architecture adopts a hierarchical feature learning design: capturing low-level morphological features in the initial layer and progressively refining them into high-level spatiotemporal feature representations in deeper layers. To achieve this, we introduce a series of bottleneck modules that significantly enhance feature expressiveness while maintaining computational efficiency through collaborative grouped convolutions and channel attention mechanisms. Details of pre-training have been described in our previous work.

For downstream task adaptation, we employ a transfer learning strategy. We retain the pre-trained backbone network parameters and replace the original classification head with a task-specific linear projection layer, whose output neuron count is aligned with the category dimension of the target task. The model's training objective is to achieve accurate classification output by minimizing the difference between the predicted distribution and the true label.

All algorithms studied are implemented using the PyTorch framework and trained on a GPU cluster consisting of $8$ NVIDIA RTX 4090 GPUs. Model optimization employed the AdamW optimizer with an initial learning rate of $1 \times 10^{-4}$ and weight decay of $1 \times 10^{-5}$. Training lasted for 30 epochs, during which the learning rate was dynamically adjusted using cosine annealing. The batch size per GPU was set to 256. To ensure the model's generalization ability and effectively mitigate overfitting, an early stopping mechanism was implemented: training was automatically terminated if the validation set loss showed no improvement for $5$ consecutive epochs.

\subsection*{Evaluation and Statistical Analysis}
Model performance was comprehensively assessed using the Area Under the Receiver Operating Characteristic curve (AUROC) to evaluate global discriminative capacity. 

To provide a granular evaluation, we further reported Sensitivity, Specificity, and the F1-score. Sensitivity quantified the model's capacity to correctly identify positive cases:
\begin{equation}
    \mathrm{Sensitivity} = \frac{\mathrm{TP}}{\mathrm{TP} + \mathrm{FN}}
\end{equation}
Specificity measured the accuracy in excluding negative controls:
\begin{equation}
    \mathrm{Specificity} = \frac{\mathrm{TN}}{\mathrm{TN} + \mathrm{FP}}
\end{equation}
The F1-score, calculated as the harmonic mean of precision and recall, served as a robust metric for overall classification accuracy, particularly in the context of class imbalance:
\begin{equation}
    F1 = 2 \cdot \frac{\mathrm{Precision} \cdot \mathrm{Recall}}{\mathrm{Precision} + \mathrm{Recall}}
\end{equation}

\section*{Results}
\subsection*{AnyECG Demonstrates Excellent Performance in Cardiovascular Disease Detection}
Here, we report the internal and external validation performance of AnyECG across a spectrum of ECG-recognizable cardiovascular diseases. Internal validation was conducted using a patient-level split of the MGH-ECG cohort, whereas the Tongji-ECG dataset was used for external testing. AnyECG achieved robust and consistent discrimination across all classical ECG-detectable abnormalities. For atrial fibrillation and flutter, AUROC reached 0.949 (95\% CI: 0.943–0.954) internally and 0.884 (95\% CI: 0.881–0.887) externally. Paroxysmal tachycardia was similarly well identified (internal AUROC: 0.852 [95\% CI: 0.839–0.868]; external AUROC: 0.841 [95\% CI: 0.836–0.846]). Conduction disorders, including atrioventricular and left bundle-branch blocks, were also identified with excellent accuracy, achieving an internal AUROC of 0.912 (95\% CI: 0.900–0.923) and an external AUROC of 0.847 (95\% CI: 0.843–0.851). Acute myocardial infarction demonstrated outstanding internal discrimination with an AUROC of 0.954 (95\% CI: 0.949–0.958), while maintaining strong generalizability externally with an AUROC of 0.850 (95\% CI: 0.845–0.858). Cardiomyopathy achieved the highest performance among all evaluated disease categories, with an internal AUROC of 0.957 (95\% CI: 0.951–0.962) and an external AUROC of 0.866 (95\% CI: 0.864–0.869).

Beyond these traditional targets, AnyECG achieved meaningful discrimination for structural and systemic cardiovascular diseases that traditionally lack definitive ECG signatures. For non-rheumatic mitral valve disease, performance reached an internal AUROC of 0.889 (95\% CI: 0.883–0.895) and an external AUROC of 0.764 (95\% CI: 0.759–0.774), whereas multi-valve involvement was identified with an internal AUROC of 0.932 (95\% CI: 0.914–0.947) and a notably strong external AUROC of 0.926 (95\% CI: 0.901–0.949). Non-rheumatic aortic valve disease yielded an internal AUROC of 0.915 (95\% CI: 0.906–0.925) and an external AUROC of 0.893 (95\% CI: 0.889–0.899), and rheumatic tricuspid valve disease achieved an internal AUROC of 0.918 (95\% CI: 0.902–0.931) with an external AUROC of 0.810 (95\% CI: 0.797–0.825).

AnyECG also demonstrated capability in detecting coronary syndromes beyond infarction: angina pectoris achieved an internal AUROC of 0.728 (95\% CI: 0.723–0.732) and an external AUROC of 0.663 (95\% CI: 0.659–0.666), while acute ischemic heart disease showed an internal AUROC of 0.814 (95\% CI: 0.808–0.819) and a strong external AUROC of 0.932 (95\% CI: 0.928–0.933). For heart failure, the model achieved an internal AUROC of 0.839 (95\% CI: 0.835–0.842) and an external AUROC of 0.783 (95\% CI: 0.780–0.785).

Hypertension and its sequelae were also identifiable: primary hypertension yielded AUROCs of 0.739 (95\% CI: 0.735–0.742) internally and 0.753 (95\% CI: 0.752–0.755) externally; hypertensive heart disease, 0.887 (95\% CI: 0.878–0.896) internally and 0.788 (95\% CI: 0.777–0.797) externally; and hypertensive chronic kidney disease, 0.759 (95\% CI: 0.718–0.793) internally and 0.705 (95\% CI: 0.701–0.711) externally. Finally, cerebral infarction was detected with internal validation of (AUROC: 0.717 [95\% CI: 0.707-0.728]) and external validation of (AUROC: 0.701 [95\% CI: 0.698-0.703]) for each respective condition.

Importantly, even for these non-classical ECG targets, internal AUROC values commonly remained in the mid-0.80s, with only modest performance attenuation (typically 0.10–0.15) in external cohorts, underscoring the broad phenotypic sensitivity and real-world generalizability of the model.

Overall, these results highlight AnyECG’s dual strengths: high-level performance on established ECG interpretations and novel detection of structural, ischemic, hypertensive, and cerebrovascular pathologies, thereby extending the ECG’s utility as a universal, noninvasive screening modality.

\subsection*{AnyECG Facilitates Comprehensive Disease Detection}

Building on AnyECG’s strong diagnostic capabilities for cardiovascular diseases, we next extended its scope to the detection of diseases across all major body systems. These include, but are not limited to, the nervous system, eye and ear disorders, the respiratory system, the genitourinary system, the digestive system, endocrine, nutritional, and metabolic diseases, neoplasms, blood and immune disorders, and the musculoskeletal system.

Figure 2 summarises performance across chapters. Median AUROC values ranged from 0.78 for psychology disease to 0.80 for circulatory diseases, with genitourinary (median 0.65) and respiratory (median 0.66) also exhibiting strong aggregate signals. In total 588 of the 1,172 phenotypes (50.2 \%) achieved AUROC $\ge$ 0.65 (Bonferroni-corrected p < 0.01); 90 phenotypes surpassed an AUROC of 0.85 (Supplementary Table A).

Endocrine \& metabolic – Disorders of mineral metabolism were predicted with exceptional accuracy (AUROC: 0.884 [95\% CI: 0.882-0.885]), while common metabolic phenotypes such as type 2 diabetes mellitus still attained an AUROC of 0.703 (95\% CI: 0.697-0.707).

Mental \& behavioural – Vascular dementia  (AUROC 0.757 [95\% CI: 0.757-0.758]) and Parkinson’s disease  (AUROC 0.761 [95\% CI: 0.723-0.797]) showed a markedly high signal, echoing emerging evidence that autonomic and conduction abnormalities accompany neurodevelopmental disorders.

Neoplastic diseases – Among Lymphoid leukemia (AUROC:  0.855 [95\% CI: 0.850-0.859]) and leukemia of unspecified cell type (AUROC: 0.798 [95\% CI: 0.792-0.802]) were the most accurately anticipated.

Respiratory system – Chronic obstructive pulmonary disease, pneumonia and respiratory failure reached AUROC values of 0.773 [95\% CI: 0.759-0.786], 0.750 [95\% CI: 0.745-0.755] and 0.740
[95\% CI: 0.730-0.750], respectively, indicating that subclinical gas-exchange or autonomic changes are already encoded in the ECG.

Digestive system – Crohn’s disease was predicted well (AUROC: 0.817 [95\% CI: 0.774-0.857]), aligning with studies linking intestinal inflammation to cardiac electrophysiology via systemic cytokine release.

Genitourinary system – Kidney failure (AUROC: 0.821 [95\% CI: 0.812-0.832]) and chronic kidney disease (AUROC: 0.765 [95\% CI: 0.759-0.771]) both surpassed clinical utility thresholds.

All highlighted phenotypes retained statistical significance after multiplicity correction. Importantly, many of the strongest signals (e.g. cardiomyopathy, disorders of mineral metabolism, lymphoid leukemia) involve pathophysiological mechanisms—myocardial remodeling, hormonal milieu, autonomic dysregulation—that plausibly manifest as subtle but learnable ECG patterns years before clinical recognition.

These results extend the utility of the resting ECG well beyond cardiovascular medicine, demonstrating that a foundation ECG model captures latent signatures of endocrine, oncological, respiratory and neuropsychiatric pathology. By illuminating hundreds of pre-symptomatic risk markers across diverse organ systems, our work positions the ECG as a scalable, low-cost screening modality for population-level precision health initiatives.

\subsection*{AnyECG Demonstrates Potential for Future Disease Prediction}

We next asked whether a single 12‑lead tracing could yield actionable prognostic information across the cardio‑kidney‑metabolic (CKM) spectrum. We evaluated AnyECG leveraging 44,336 ECGs from the CKB. For each of the eight most prevalent CKM conditions we allocated participants to low‑, medium‑, or high‑risk tertiles and evaluated incident disease–free survival over a median follow‑up of 8 years.

In all CKM-related diseases evaluated, AnyECG demonstrated clear and statistically significant separation of Kaplan–Meier survival curves (log-rank p < 0.001 for all comparisons; Figures 4a–d and 5a–d). For example, in essential hypertension (Fig. 4a), consistent with clinical expectations, the 10-year disease-free survival rate was markedly lower in the high-risk group than in the low-risk group (83.7\% vs. 94.4\%), with a hazard ratio (HR) of 2.08 (95\% CI 1.95–2.22). Acute myocardial infarction (Fig. 4b) exhibited a similarly pronounced difference (97.2\% vs. 99.4\%; HR = 2.41, 95\% CI 2.15–2.70). Atrial fibrillation (Fig. 4c) showed a 10-year survival of 92.0\% versus 98.0\% between the high- and low-risk groups (HR = 2.46, 95\% CI 2.28–2.65), and heart failure (Fig. 4d) displayed a comparable separation (95.6\% vs. 99.0\%; HR = 2.73, 95\% CI 2.43–3.07).

Systemic CKM diseases demonstrated similarly robust risk stratification. Cerebral infarction (Fig. 5a) showed survival rates of 97.8\% versus 99.5\% (HR = 2.18, 95\% CI 1.96–2.42), type 2 diabetes mellitus (Fig. 5b) 95.1\% versus 98.6\% (HR = 2.29, 95\% CI 2.12–2.47), chronic kidney disease (Fig. 5c) 96.3\% versus 98.9\% (HR = 2.11, 95\% CI 1.96–2.27), and gout (Fig. 5d) 96.9\% versus 99.1\% (HR = 2.05, 95\% CI 1.88–2.24).

Together, these results highlight that a single resting ECG, interpreted by AnyECG, captures latent systemic health signals associated with CKM disease development, enabling clinically actionable and scalable population-level risk stratification.

Even for diseases with extremely low absolute incidence, such as acute myocardial infarction and cerebral infarction, AnyECG still produced narrow confidence intervals and significant discrimination, highlighting the statistical robustness and practical predictive power of the model-derived biomarkers. In summary, a single resting electrocardiogram interpreted by AnyECG can capture the implicit characteristics of systemic health status, and its applicability extends beyond the traditional risk prediction paradigm centered on cardiac endpoints. This method provides a simultaneously available, multi-indicator consistent risk assessment capability for common chronic diseases, laying an important foundation for low-cost, non-invasive, scalable population risk stratification and precision prevention strategies.

\subsection*{AnyECG Reveals Links Between Comorbidities}

We quantified pair‑wise associations between all model‑derived risk scores in the 220,187 Tongji-ECG participants and visualised the results with complementary network, scatter‑plot and heat‑map representations (Fig. 6).

The rank‑correlation heat‑map of all diseases (Fig. 6a) displayed a striking block‑diagonal architecture. A highly cohesive cardiometabolic module—comprising essential hypertension (HPT), atrial fibrillation (AF), myocardial infarction (I21), cerebral infarction (I64) and rheumatic heart disease (RHD)—showed a median Spearman r = 0.42 (inter‑quartile range, IQR 0.38–0.48). Pulmonary (COPD), kidney (CKD) and metabolic (type 2 diabetes mellitus, T2DM; gout) conditions clustered separately (median intra‑cluster r = 0.28), whereas neuro‑degenerative and anaemia‑related codes (Parkinson’s disease, PD; vascular dementia, VaD; iron‑deficiency anaemia, IDA) formed a comparatively weakly connected community (median r = 0.07). These patterns mirror epidemiological studies, suggesting that AnyECG eyes the same latent pathophysiology directly from electrical activity.

The chord diagram for the 15 most common conditions (Fig. 6b) translated the correlation matrix into an interpretable network: link width encodes mutual information and colour denotes direction (red positive, blue negative). Hypertension emerged as the most central node (weighted degree = 33.7), radiating strong connections to AF, T2DM and CKD, followed by AF (degree = 31.2) and CKD (degree = 29.5). Disease‑specific subnetworks underscore clinically recognised cascades:

Individual‑level scatter plots confirm selective, not global, correlations.
To exclude the possibility that AnyECG simply assigns uniformly high risk to sicker individuals, we inspected representative bivariate distributions. Predicted stroke (I64) and myocardial infarction (I21) scores were tightly co‑distributed (Spearman r = 0.67; Fig. 6g), whereas Parkinson’s disease (G30) and stroke scores were almost uncorrelated (r = 0.05; Fig. 6h). Similarly, kidney disease (N18) and ischaemic bowel disease (I25) predictions showed no appreciable structure (r = 0.02; Fig. 6i). Thus, the comorbidity signal is specific, not an artefact of a generic “bad ECG” phenotype.

These analyses demonstrate that AnyECG does not treat diseases as isolated endpoints; instead, it learns a latent representation that encodes the complex topology of human multimorbidity. By jointly modelling cross‑organ interactions, an inexpensive resting ECG can therefore serve as a non‑invasive window into an individual’s integrated disease trajectory—a capability with immediate relevance for differential diagnosis, risk‑aligned screening programmes, and cluster‑targeted precision prevention.

\begin{figure}[]
    \centering
    \includegraphics[width=\textwidth]{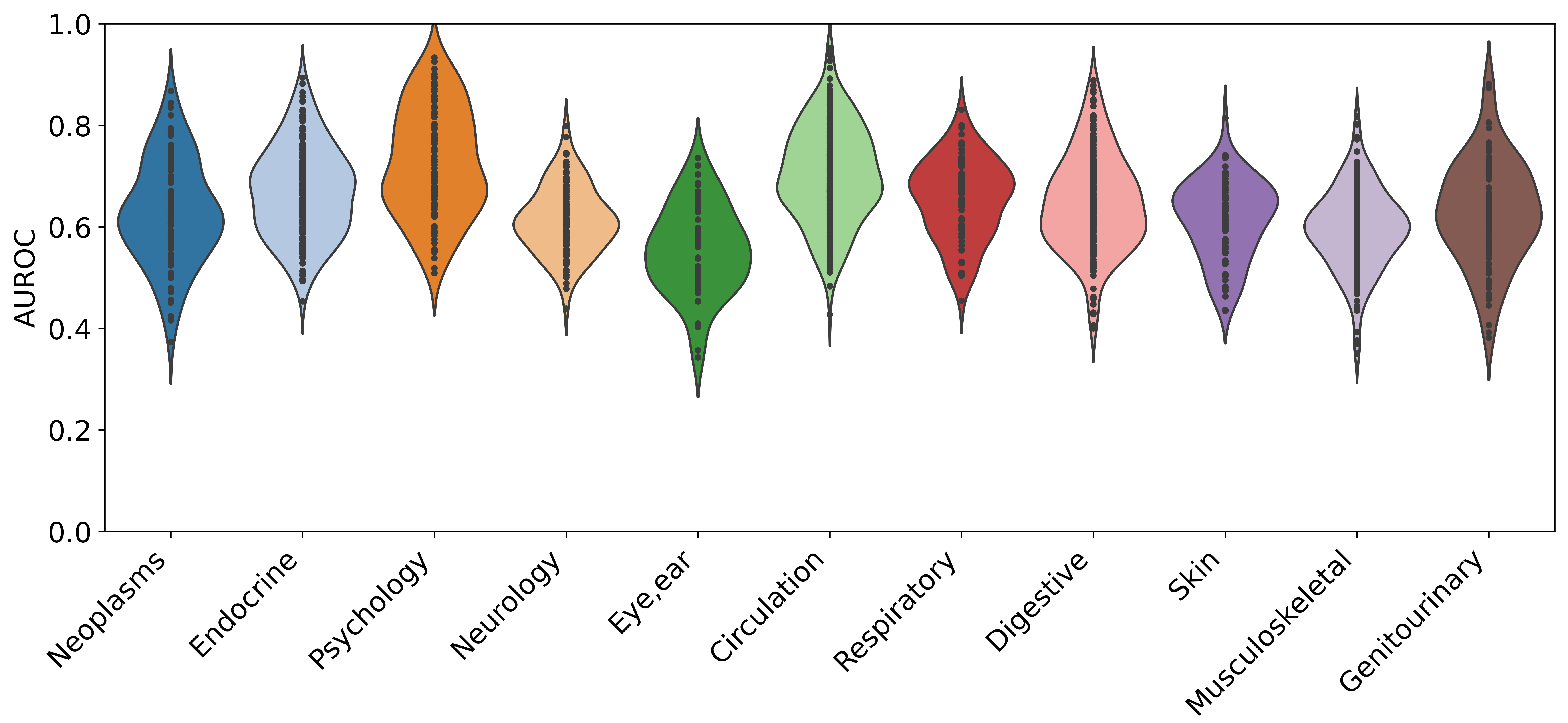} 
    \caption{Boxplots for the overall internal test set performance of AnyECG on different disease systems.}
    \label{fig:boxplot}
    \vspace{2mm}
    \hrule
\end{figure}

\begin{figure}[]
    \centering
    \includegraphics[width=\textwidth]{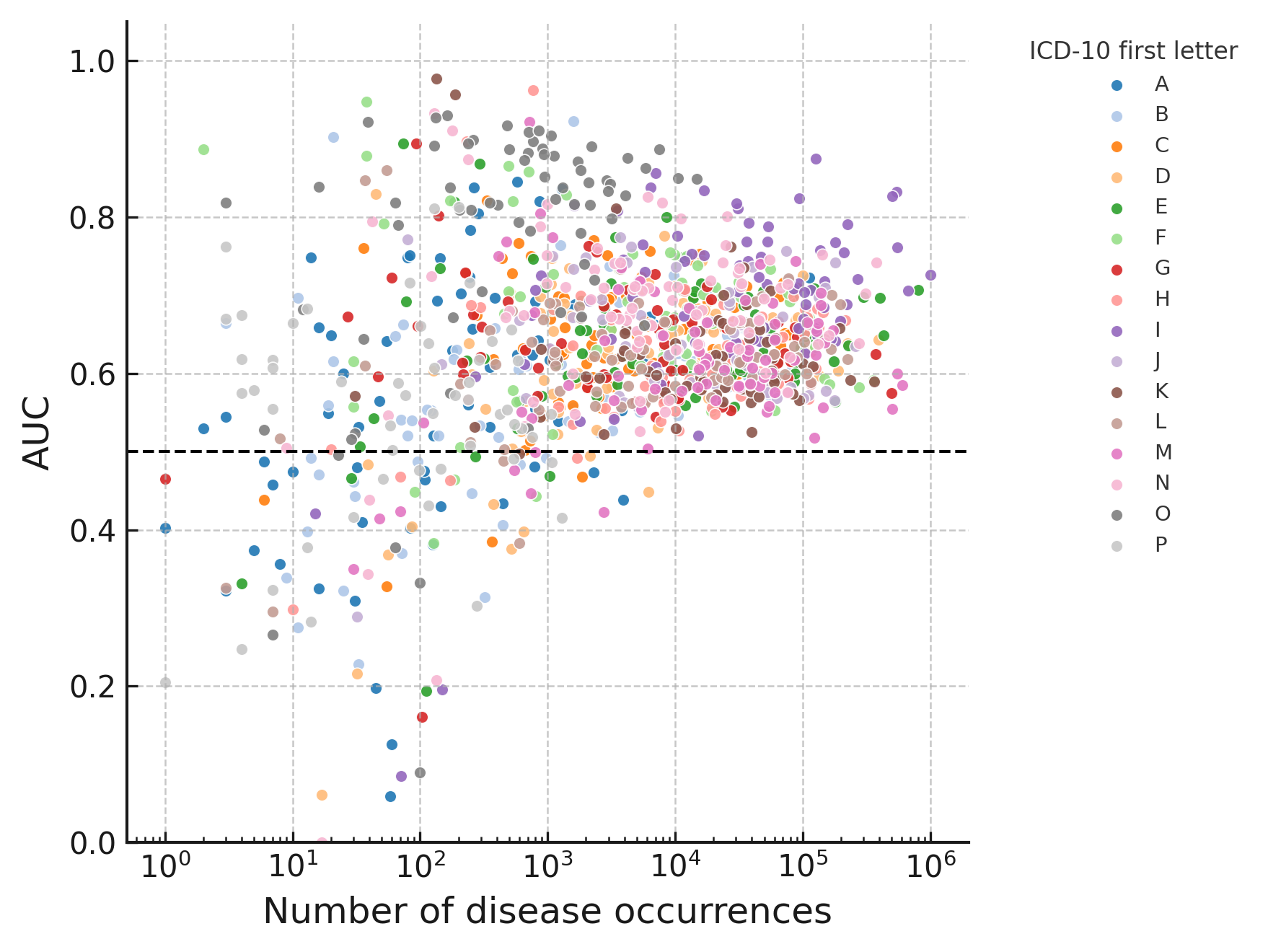} 
    \caption{Average diagnosis AUC values (y axis) as a function of training occurrences (x axis).}
    \label{fig:disease}
    \vspace{2mm}
    \hrule
\end{figure}

\begin{figure*}[!t]
\centering
\begin{subfigure}{0.45\textwidth}
    \includegraphics[width=\textwidth]{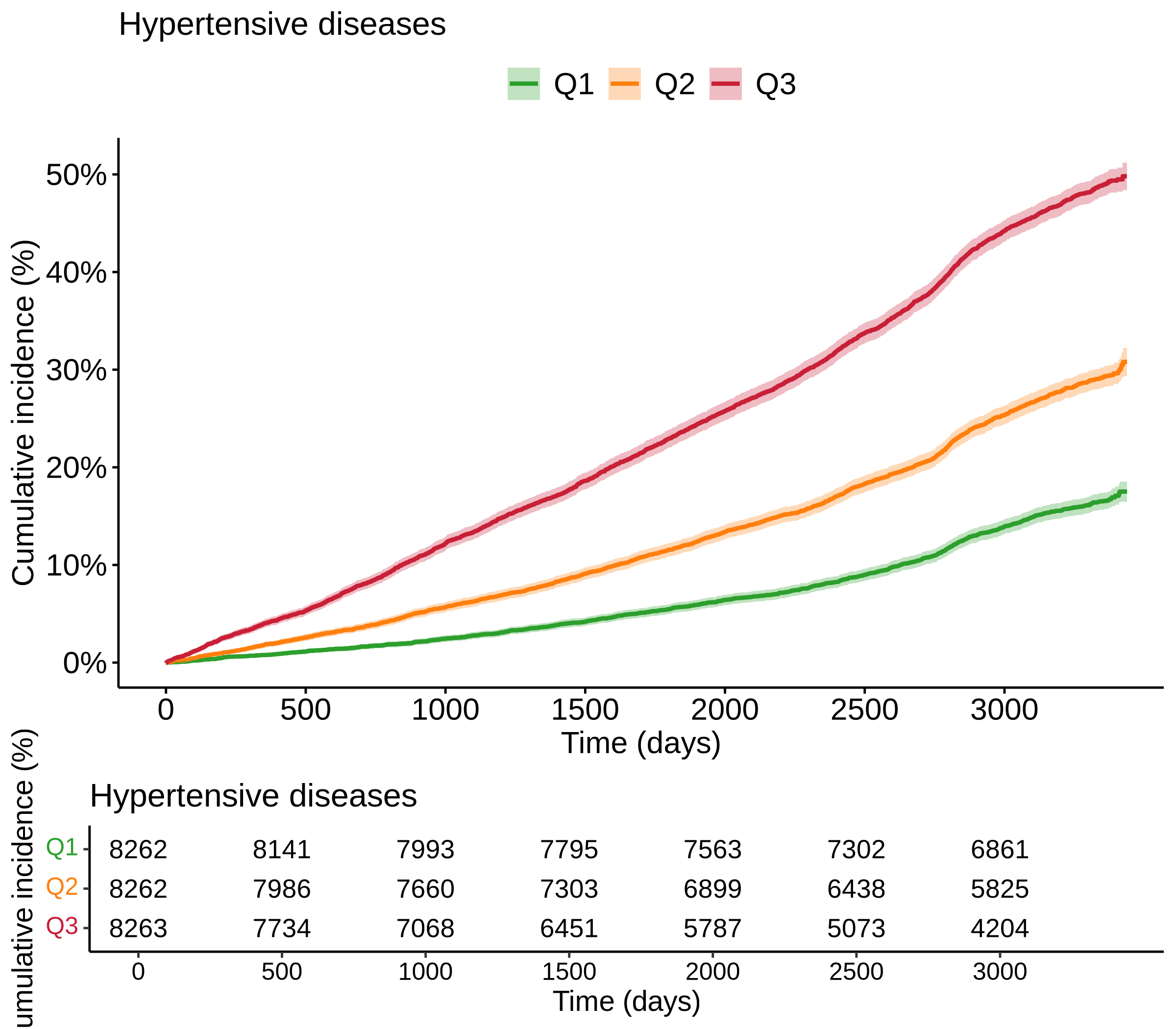}
    \caption{Hypertensive diseases}
\end{subfigure}
\hfill
\begin{subfigure}{0.45\textwidth}
    \includegraphics[width=\textwidth]{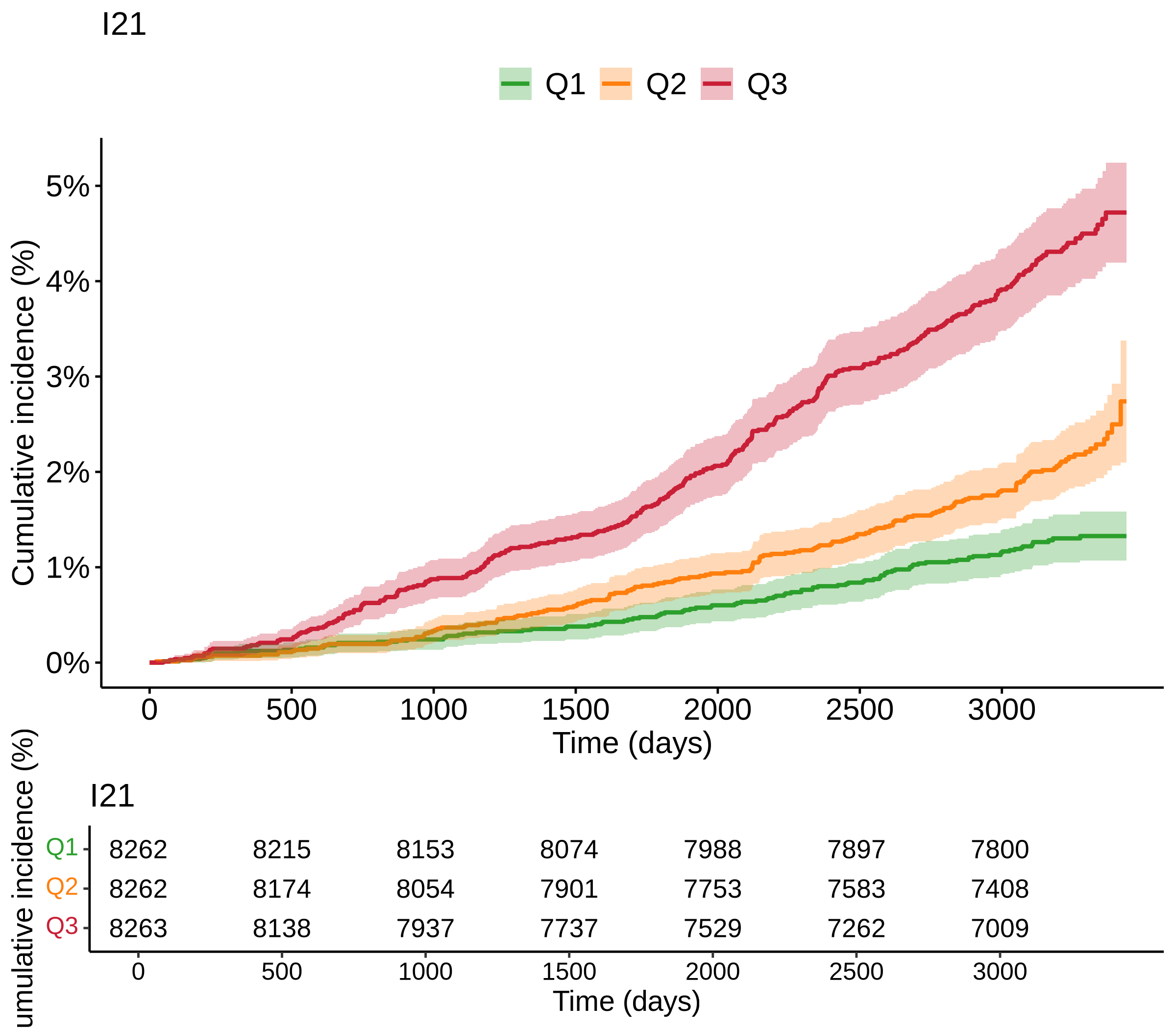}
    \caption{Acute myocardial infarction}
\end{subfigure}

\vspace{3mm}

\begin{subfigure}{0.45\textwidth}
    \includegraphics[width=\textwidth]{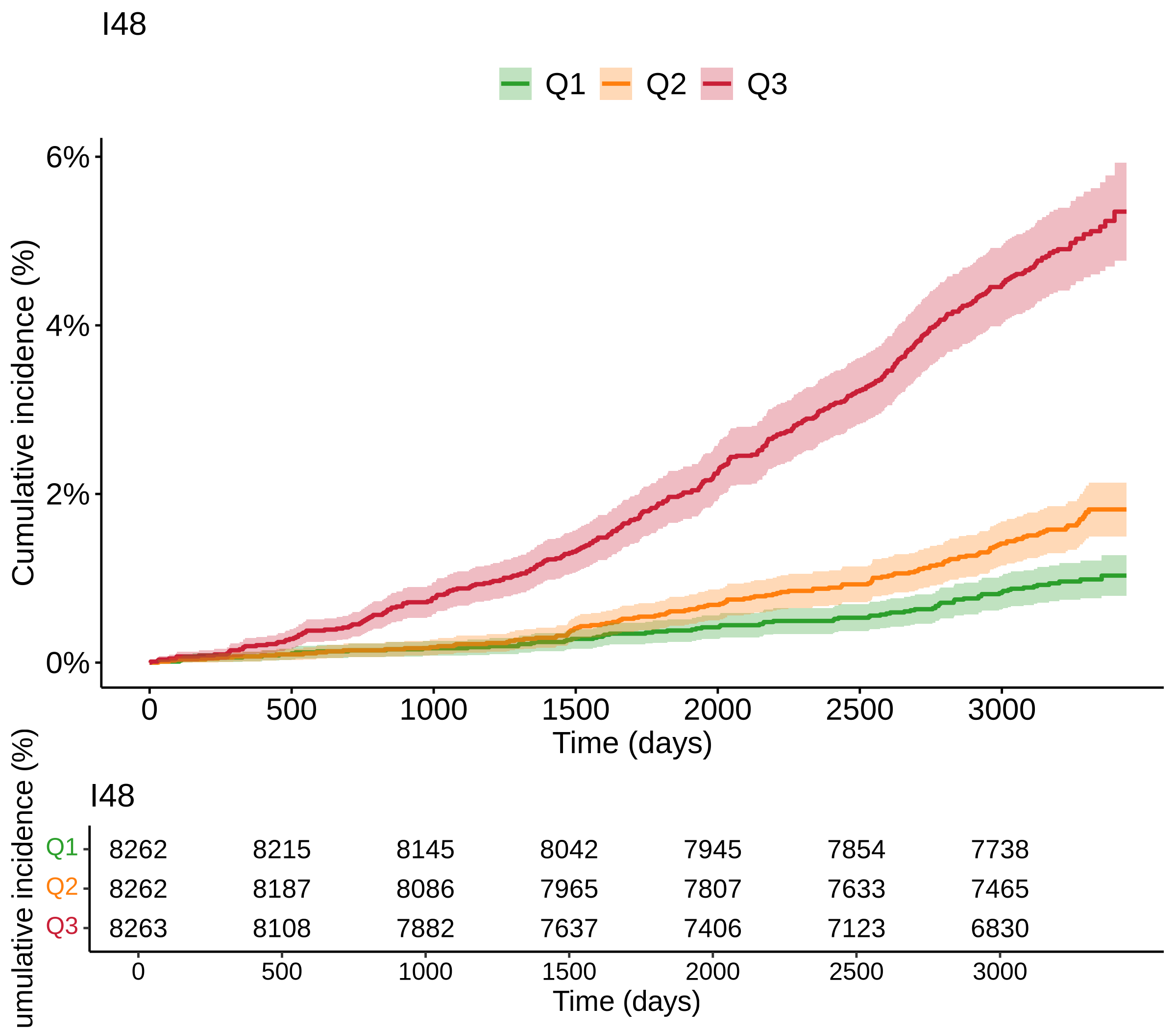}
    \caption{Atrial fibrillation and flutter}
\end{subfigure}
\hfill
\begin{subfigure}{0.45\textwidth}
    \includegraphics[width=\textwidth]{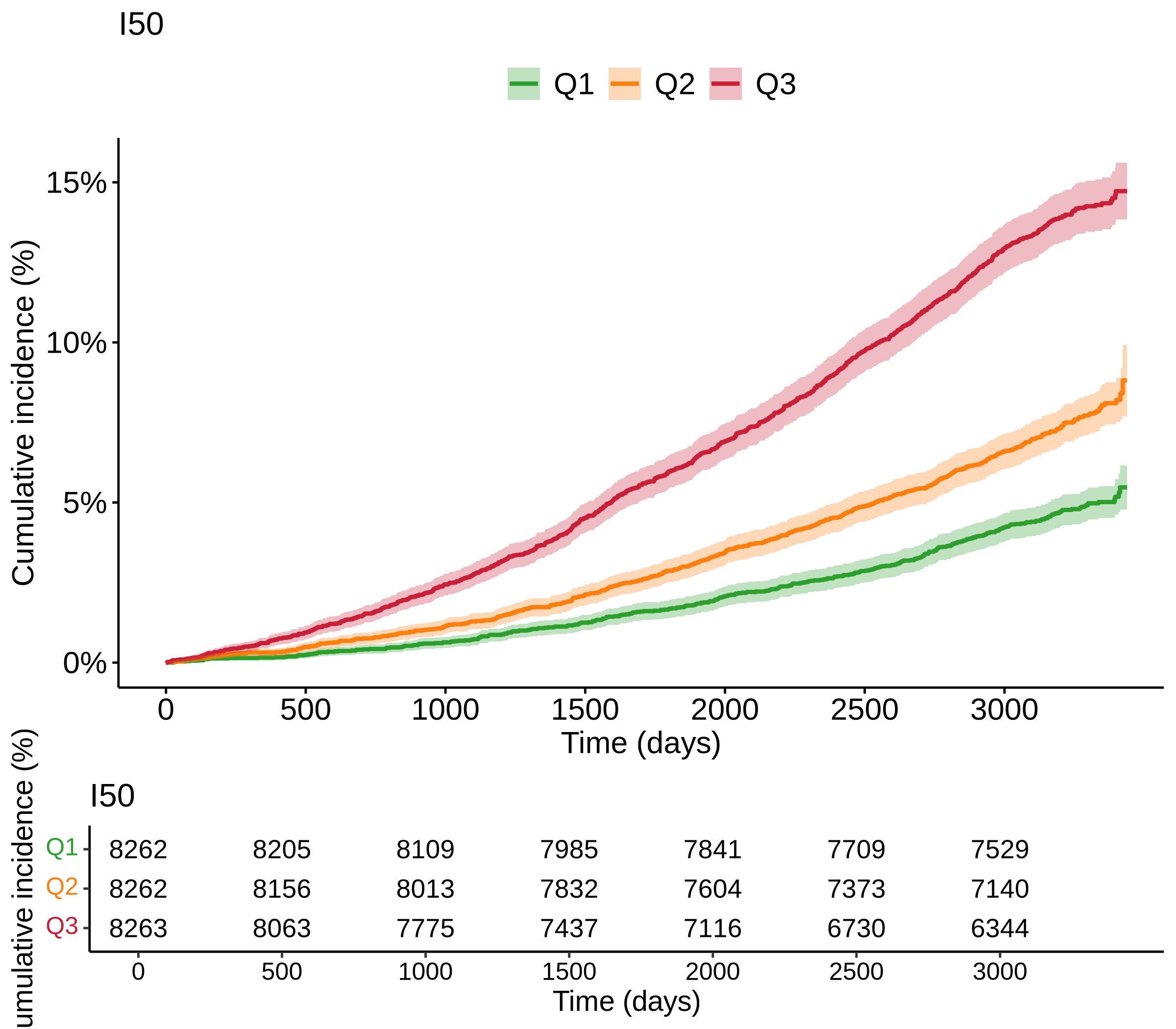}
    \caption{Heart failure}
\end{subfigure}

\caption{\textbf{Prognostic stratification of major cardiometabolic cardiac events using AnyECG in the CKB cohort.}}
\label{fig:ckb_cardiac}
\vspace{2mm}
\hrule
\end{figure*}

\begin{figure*}[!t]
\centering
\begin{subfigure}{0.45\textwidth}
    \includegraphics[width=\textwidth]{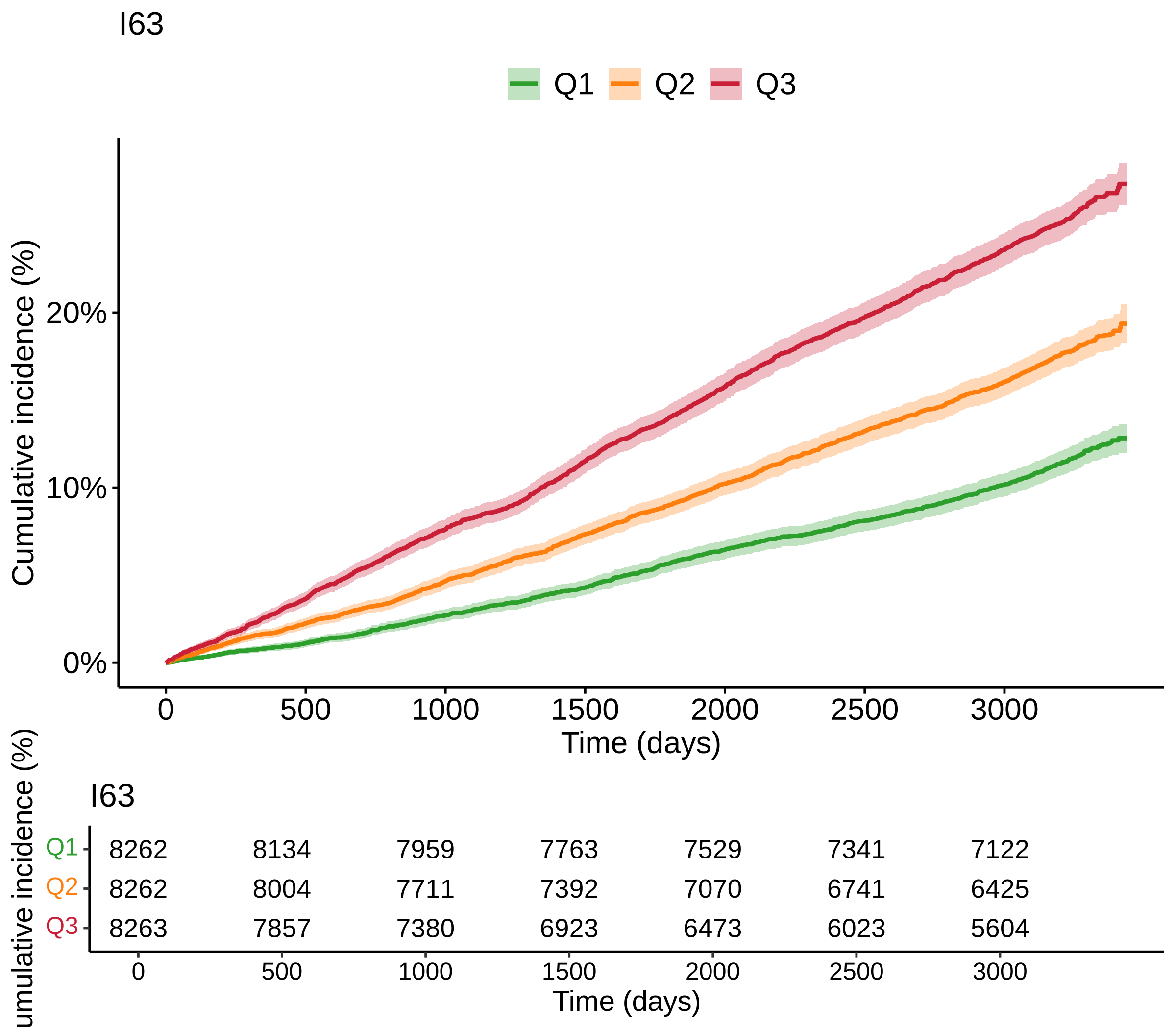}
    \caption{Cerebral infarction}
\end{subfigure}
\hfill
\begin{subfigure}{0.45\textwidth}
    \includegraphics[width=\textwidth]{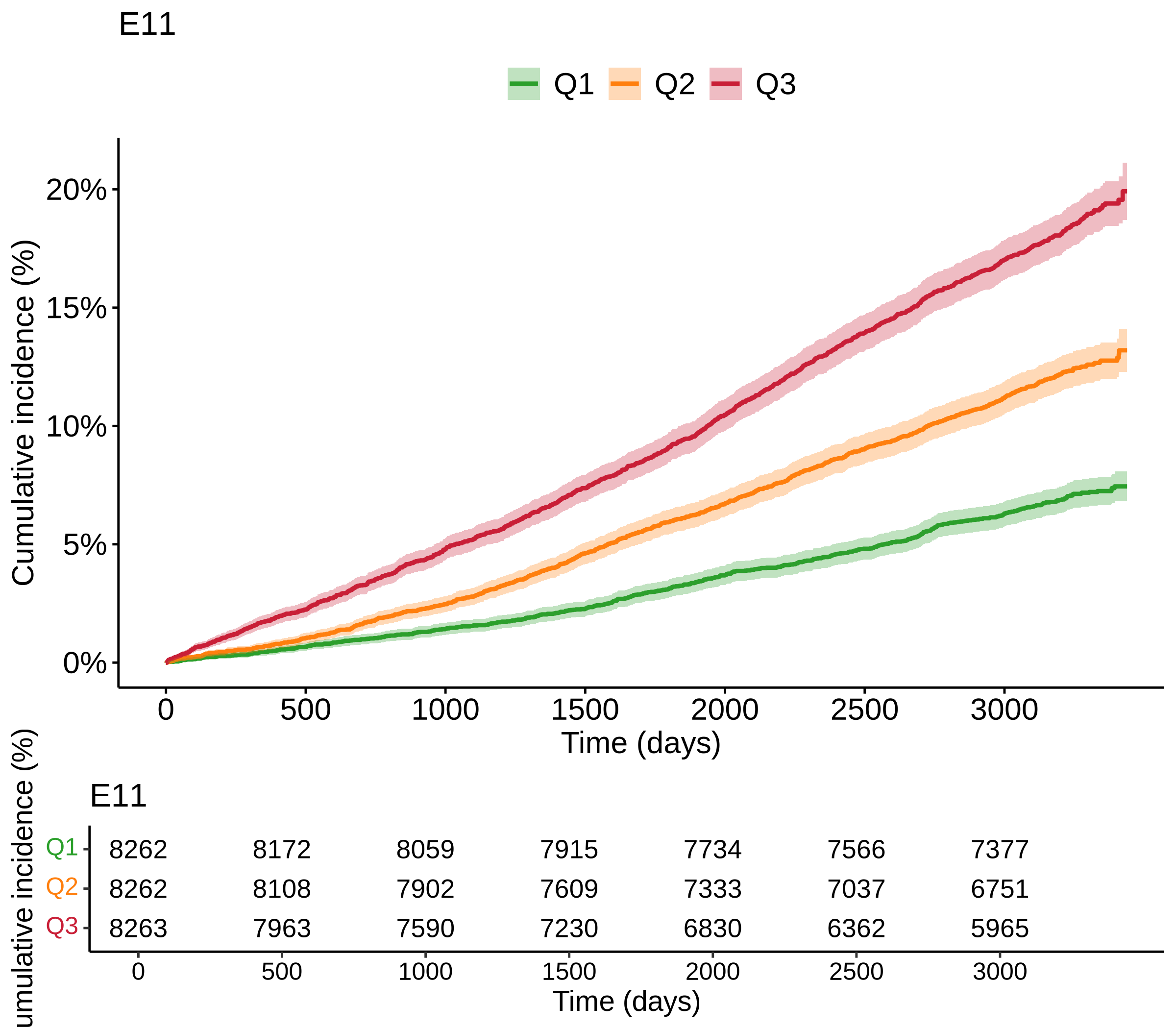}
    \caption{Type 2 diabetes}
\end{subfigure}

\vspace{3mm}

\begin{subfigure}{0.45\textwidth}
    \includegraphics[width=\textwidth]{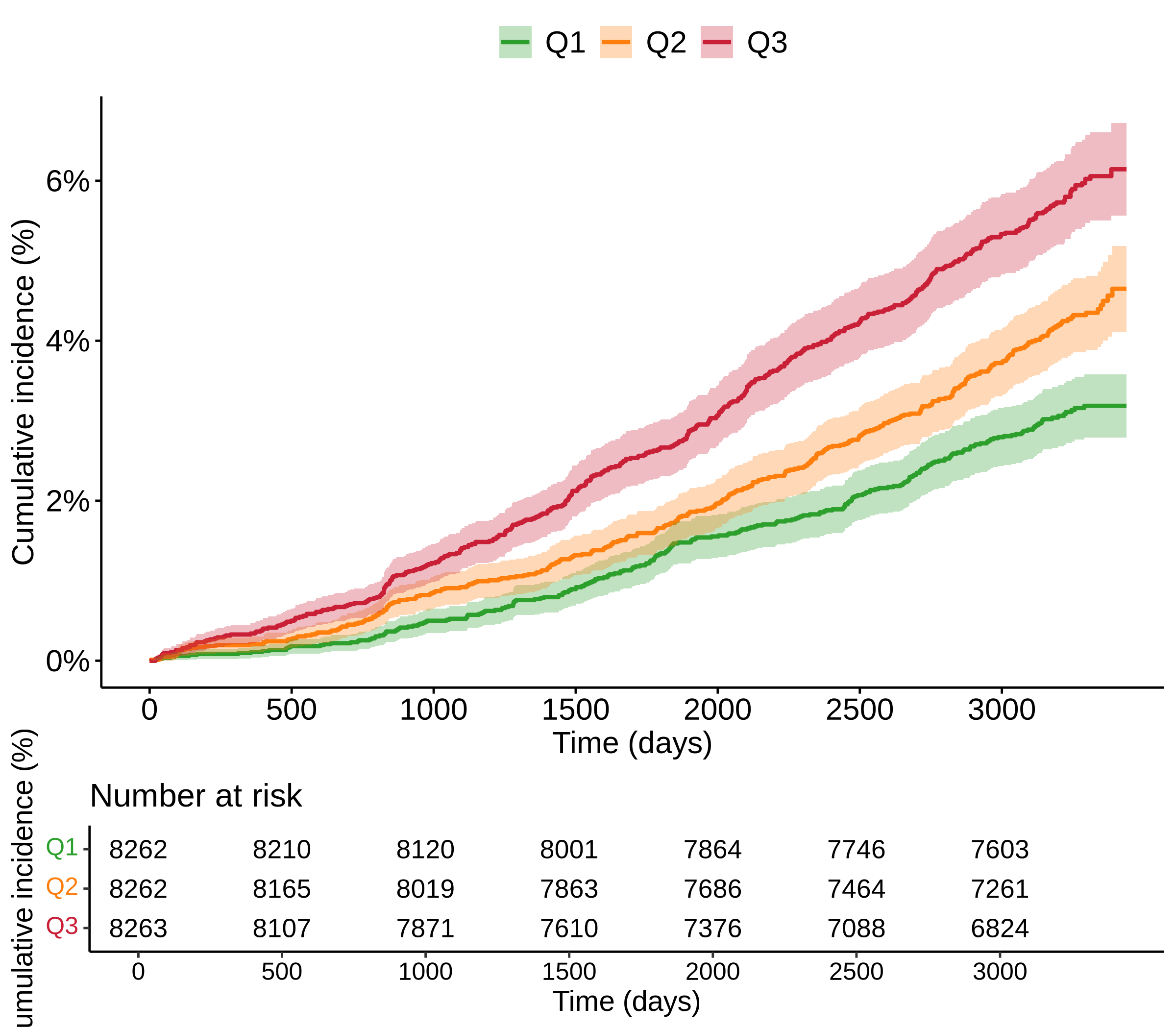}
    \caption{Chronic kidney disease}
\end{subfigure}
\hfill
\begin{subfigure}{0.45\textwidth}
    \includegraphics[width=\textwidth]{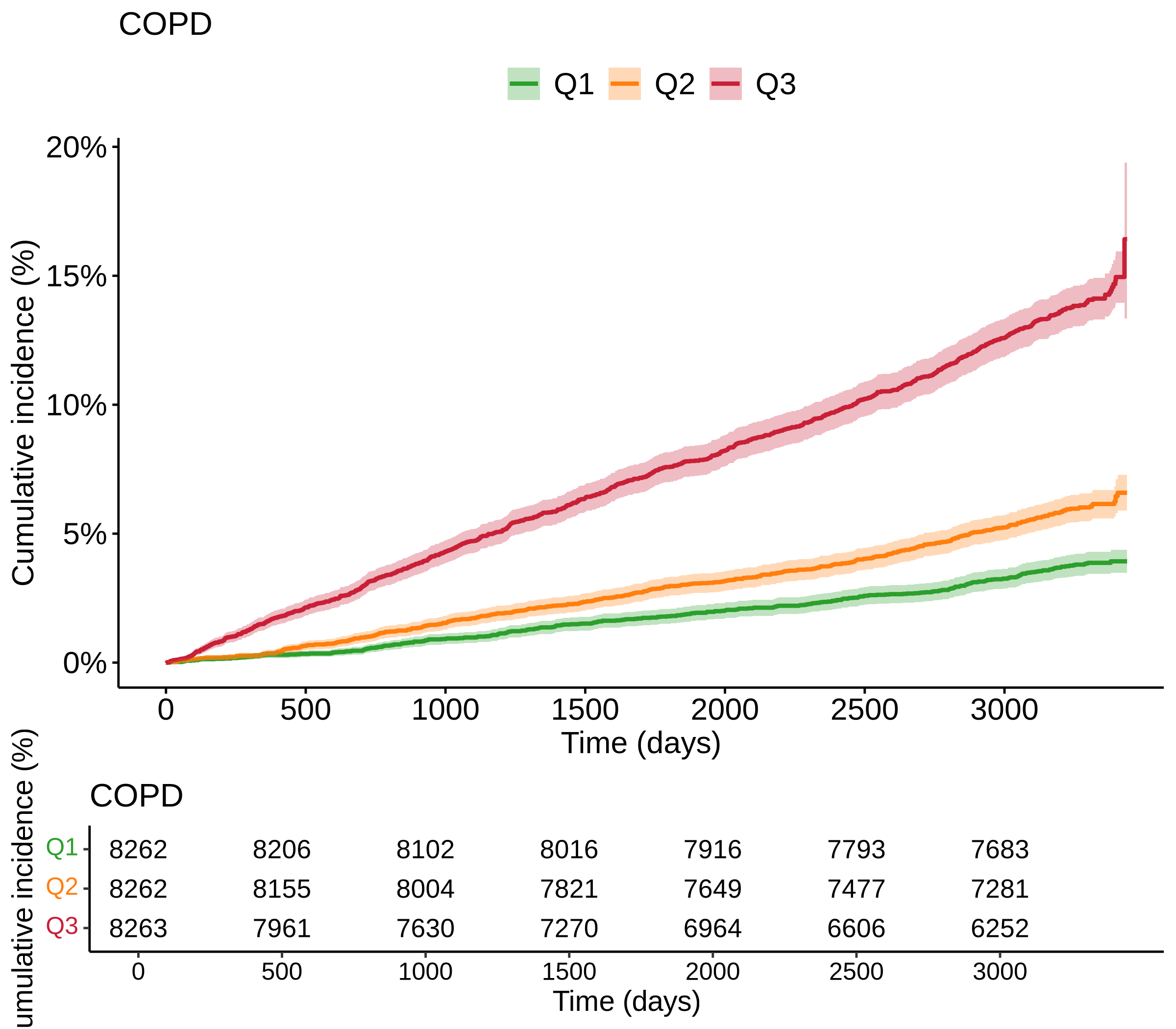}
    \caption{Gout}
\end{subfigure}

\caption{\textbf{Prognostic stratification of systemic CKM diseases using AnyECG in the CKB cohort.}}
\label{fig:ckb_ckm}
\vspace{2mm}
\hrule
\end{figure*}

\begin{figure}[]
    \centering
    \begin{subfigure}[t]{0.98\textwidth}
   \centering
   \includegraphics[width=\linewidth]{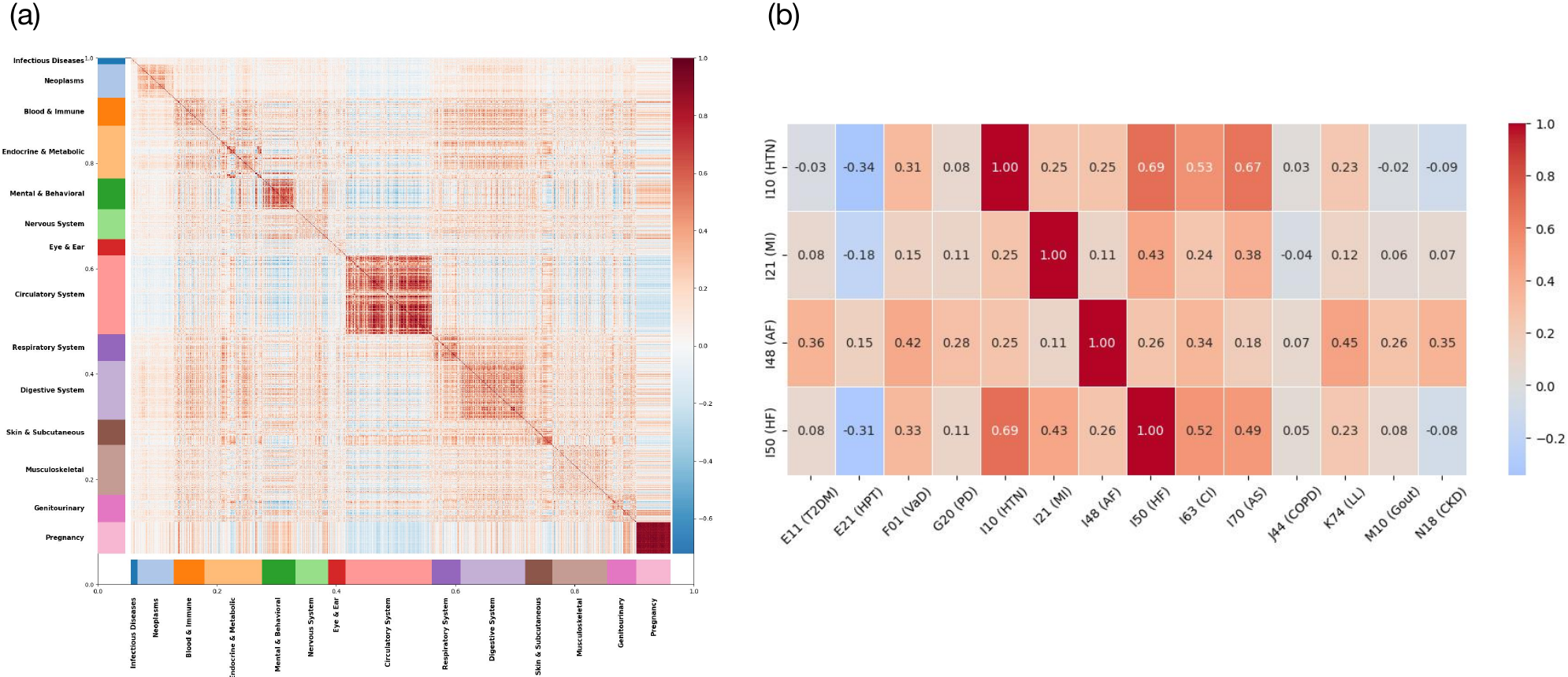}
    \end{subfigure}
    \caption{Model-derived disease co-occurrence patterns and correlation structure. (A) Chord diagram showing pairwise associations between selected chronic diseases, where ribbon width reflects the strength of association and ribbon color encodes the sign and magnitude (red: stronger positive association; blue: negative/weaker association). Disease segments on the circumference are colored by clinical system (e.g., cardiovascular, metabolic, kidney, respiratory). (B) Correlation heatmap between columns of the model output, illustrating the global correlation structure across all predicted disease labels and revealing blocks of tightly correlated conditions along the diagonal.}
    \vspace{2mm}
    \hrule
\end{figure}


\vspace{1em}










\section*{Discussion}



In our study, we have developed a comprehensive hybrid model for diagnosing various diseases across different human systems, using the electrocardiogram (ECG) as the sole input. Utilizing the Tongji-ECG dataset as the training set and a large Dataset from prominent tertiary hospitals in China as an external validation set, we have doubly confirmed the potential of ECG in diagnosing a broad spectrum of diseases. This provides a scientific basis for future single-modal, non-invasive, and early full-spectrum diagnoses using ECG alone.

\subsection*{Revolutionizing ECG as a Systemic Diagnostic Tool}
The AnyECG model represents a paradigm shift in electrocardiography, demonstrating unprecedented capacity to detect latent cardiovascular disease (CVD) and diverse non-cardiac conditions years before clinical manifestation. Specifically, ECG as a standard way in recording the electrical activity of heart, shows specific performance in detecting abnormalities in cardiovascular diseases. In our study, its exceptional performance for cardiomyopathy (AUROC 0.95), AMI (AUROC 0.95), structural heart diseases (AUROC 0.94), and atrial fibrillation (AUROC 0.94) aligns with known electrophysiological substrates but extends beyond current clinical expectations, indicating the consistent and excellent monitoring performance of AI-enabled model in a wide spectrum of cardiovascular disease system. More remarkably, the model identified high-risk signatures for conditions spanning endocrine (puberty disorders, AUROC 0.98), neuropsychiatric (ADHD, AUROC 0.99), oncologic (breast cancer, AUROC 0.83), and gastrointestinal diseases (Crohn’s, AUROC 0.82). This evidences that ECG morphology encodes systemic pathophysiology—likely mediated through autonomic nervous system dysregulation, inflammatory cascades, or metabolic stress—transforming the ECG from a cardiac-specific tool into a holistic biomarker platform. 
For example, Sunil Vasu Kalmady developed a machine learning algorithms based on electrocardiograms for cardiovascular diagnoses at the population level.\cite{kalmady2024development} Nils Strodthoff et.al. construct an explorative study in emergency care using AI-enhanced ECG as a screening tool for cardiac and non-cardiac conditions.\cite{strodthoff2024prospects}  Sam F. Friedman analyzed ECG encodings and different diseases Phecodes using AI model and found that the strongest ECG association was with hypertension.\cite{friedman2025unsupervised} Building on the aforementioned studies, our research has taken a further step by conducting a more comprehensive and broad-spectrum analysis of the end-to-end diagnostic links between ECG and individual diagnoses across various systems.

\subsection*{ECG as Novel Digital Biomarkers for Systemic Diseases to Explore the Mechanistic Plausibility and Biological Insights}

The observed discriminative power finds strong biological grounding in shared subclinical pathways. Conditions with peak accuracy—cardiomyopathy, ADHD, and puberty disorders involve autonomic remodeling or hormonal modulation known to alter repolarization dynamics and conduction stability.\cite{sekaninova2020role} Core symptoms of ADHD (such as attention deficit and hyperactivity) are closely associated with autonomic nervous dysfunction (e.g., sympathetic hyperactivity), suggesting that similar mechanisms may underlie the link between neurodevelopmental disorders and electrophysiological abnormalities.\cite{devine2021paraneoplastic} Hormonal dysregulation (e.g., thyroid hormones and sex hormones) can directly influence repolarization dynamics and conduction stability by altering the expression of myocardial ion channels.\cite{amin2010cardiac,cao2025glycolytic} Patients with puberty disorders often exhibit hormonal fluctuations, which may induce electrophysiological remodeling analogous to paraneoplastic arrhythmias, manifesting as QT interval prolongation or premature ventricular contractions.\cite{d2017electrocardiographic} Similarly, inflammatory states (e.g., Crohn’s disease) may drive electrophysiological shifts via cytokine-mediated ion channel dysfunction,\cite{stringer2023pathophysiology,werlein2023inflammation} while metabolic disorders like obesity (AUROC 0.90) likely manifest through myocardial substrate utilization defects. For oncology, the prediction of breast and thyroid cancers years before diagnosis supports emerging literature on paraneoplastic electrophysiological remodeling.\cite{iorio2019paraneoplastic,soomro2020paraneoplastic} Ben G T Coumbe et.al. reports electrophysiological abnormalities resulting from autonomic nervous dysfunction in patients with breast cancer and other malignancies, and cardiac autonomic nervous dysfunction in cancer patients encompasses electrophysiological abnormalities such as paraneoplastic arrhythmias.\cite{coumbe2018cardiovascular} These findings suggest ECG signatures reflect not only structural cardiac changes but also neuroendocrine, inflammatory, and metabolic cross-talk, enabling early risk stratification across organ systems.

\subsection*{Endow ECG with New Applications in Detecting  Comorbidities and Predicting Future Diseases}

In our study, we found that there are close associations between different system diseases, as well as in different diseases in one system, indicating the ECG foundation model has endowed ECG new applications for diagnosing comorbid diseases. For example, Hypertension is one of the most common and potent risk factors for developing atrial fibrillation.\cite{kamioka2024hypertension} While chronically high blood pressure forces the heart to work harder causing Left Ventricular Hypertrophy (LVH), left Atrial Enlargement, fibrosis and Electrical Remodeling, which serves as the reason of AF.\cite{rohla2025summary} Besides, HTN often coexists with other AF risk factors including coronary artery disease (CAD) and heart failure (HF), which strengthen the association of AF and HTN.\cite{potpara20242024} Further, Uncontrolled hypertension damages the delicate blood vessels (glomeruli) in the kidneys over time, reducing their filtering ability.\cite{suarez2025hypertension} At the meantime, CKD induced deficient ability to regulate fluid and electrolyte balance, and thus activate the Renin-Angiotensin-Aldosterone System (RAAS), and impair blood vessel function, which leads to progressive HTN.\cite{halimi2024management} 

Further, we found potential interactions between COPD and RHD, where there seems no direct causal link like the previous pairs. The interaction mechanism may origins from Pulmonary Hypertension (PH). Severe COPD can lead to PH, and increase the pressure in the right side of the heart.\cite{mullen2025recent} If RHD has caused right-sided valve disease (like tricuspid stenosis/regurgitation), this pre-existing damage is significantly worsened by the increased workload from PH.\cite{rwebembera2024acute} Besides, hypoxia and Inflammation in COPD can exacerbate underlying heart disease including RHD.\cite{leitner2025update,stieglitz2025copd} Thus the ECG-foundation model not only shows consisted mechanisms, but also exhibites potential and latent mechanism compared to the previous studies. This indicates it a valuable tool in diseases treatment and prognosis prediction.

\section*{Conclusion}

In conclusion, this study fundamentally transforms the diagnostic potential of the electrocardiogram (ECG). Through the development and rigorous external validation of the AnyECG model, we demonstrate that AI-powered analysis of a single, standard ECG can provide highly accurate diagnosis (as evidenced by high AUROCs) not only for a wide spectrum of cardiovascular diseases (e.g., cardiomyopathy, AMI, AF) but also for diverse non-cardiac conditions across endocrine, neuropsychiatric, oncologic, and gastrointestinal systems. This unprecedented systemic diagnostic capability is mechanistically plausible, grounded in ECG morphology capturing systemic pathophysiology mediated by autonomic nervous system dysregulation, hormonal modulation, inflammatory cascades, metabolic stress, and paraneoplastic effects and so on. Furthermore, the ECG foundation model reveals intricate associations and potential pathways linking comorbid diseases (e.g., HTN/AF, HTN/CKD, COPD/RHD via PH), offering valuable insights for early risk stratification, diagnosing complex comorbidities, and predicting long-term prognosis. Collectively, this research establishes the ECG as a powerful, non-invasive platform for holistic, early, and broad-spectrum disease detection across multiple organ systems.

\section*{ACKNOWLEDGMENTS}


This work is supported by the National Natural Science Foundation of China (62102008, 62172018), CCF-Tencent Rhino-Bird Open Research Fund (CCF-Tencent RAGR20250108), CCF-Zhipu Large Model Innovation Fund (CCF-Zhipu202414), PKU-OPPO Fund (BO202301, BO202503), Research Project of Peking University in the State Key Laboratory of Vascular Homeostasis and Remodeling (2025-SKLVHR-YCTS-02). Hongling Zhu is funded by the National Key R\&D Program of China (No. 2022YFE0209900), the National Natural Science Foundation of China (No. 82100531), Medical Artificial Intelligence Fund of Tongji Hospital (No. AI2024A02), Tongji Hospital Excellent Young Scientist Fund (No. 24-2KYC13057-14), Huazhong University of Science and Technology's 2024 "Interdisciplinary Research Support Program" Project (No. 2024JCYJ066-PP, 5003540166) and Tongji Hospital Medical Innovation and Translational Incubation Project (No. 2023CXZH004) .

\newpage

\newpage


\bibliography{reference}

@article{gow2023mimic,
  title={Mimic-iv-ecg-diagnostic electrocardiogram matched subset},
  author={Gow, Brian and Pollard, Tom and Nathanson, Larry A and Johnson, Alistair and Moody, Benjamin and Fernandes, Chrystinne and Greenbaum, Nathaniel and Berkowitz, Seth and Moukheiber, Dana and Eslami, Parastou and others},
  journal={Type: dataset},
  year={2023}
}

@article{liu2021deep,
  title={Deep learning in ECG diagnosis: A review},
  author={Liu, Xinwen and Wang, Huan and Li, Zongjin and Qin, Lang},
  journal={Knowledge-Based Systems},
  volume={227},
  pages={107187},
  year={2021},
  publisher={Elsevier}
}

@inproceedings{hong2020holmes,
  title={HOLMES: Health OnLine Model Ensemble Serving for Deep Learning Models in Intensive Care Units},
  author={Hong, Shenda and Xu, Yanbo and Khare, Alind and Priambada, Satria and Maher, Kevin and Aljiffry, Alaa and Sun, Jimeng and Tumanov, Alexey},
  booktitle={Proceedings of the 26th ACM SIGKDD International Conference on Knowledge Discovery \& Data Mining},
  pages={1614--1624},
  year={2020}
}

@article{kalmady2024development,
  title={Development and validation of machine learning algorithms based on electrocardiograms for cardiovascular diagnoses at the population level},
  author={Kalmady, Sunil Vasu and Salimi, Amir and Sun, Weijie and Sepehrvand, Nariman and Nademi, Yousef and Bainey, Kevin and Ezekowitz, Justin and Hindle, Abram and McAlister, Finlay and Greiner, Russel and others},
  journal={npj Digital Medicine},
  volume={7},
  number={1},
  pages={133},
  year={2024},
  publisher={Nature Publishing Group UK London}
}

@article{chen2011china,
  title={China Kadoorie Biobank of 0.5 million people: survey methods, baseline characteristics and long-term follow-up},
  author={Chen, Zhengming and Chen, Junshi and Collins, Rory and Guo, Yu and Peto, Richard and Wu, Fan and Li, Liming},
  journal={International journal of epidemiology},
  volume={40},
  number={6},
  pages={1652--1666},
  year={2011},
  publisher={Oxford University Press}
}

@article{strodthoff2024prospects,
  title={Prospects for artificial intelligence-enhanced electrocardiogram as a unified screening tool for cardiac and non-cardiac conditions: an explorative study in emergency care},
  author={Strodthoff, Nils and Lopez Alcaraz, Juan Miguel and Haverkamp, Wilhelm},
  journal={European Heart Journal-Digital Health},
  volume={5},
  number={4},
  pages={454--460},
  year={2024},
  publisher={Oxford University Press UK}
}

@article{friedman2025unsupervised,
  title={Unsupervised deep learning of electrocardiograms enables scalable human disease profiling},
  author={Friedman, Sam F and Khurshid, Shaan and Venn, Rachael A and Wang, Xin and Diamant, Nate and Di Achille, Paolo and Weng, Lu-Chen and Choi, Seung Hoan and Reeder, Christopher and Pirruccello, James P and others},
  journal={npj Digital Medicine},
  volume={8},
  number={1},
  pages={23},
  year={2025},
  publisher={Nature Publishing Group UK London}
}

@article{sekaninova2020role,
  title={Role of neuroendocrine, immune, and autonomic nervous system in anorexia nervosa-linked cardiovascular diseases},
  author={Sekaninova, Nikola and Bona Olexova, Lucia and Visnovcova, Zuzana and Ondrejka, Igor and Tonhajzerova, Ingrid},
  journal={International Journal of Molecular Sciences},
  volume={21},
  number={19},
  pages={7302},
  year={2020},
  publisher={MDPI}
}

@article{devine2021paraneoplastic,
  title={Paraneoplastic neurological syndromes: clinical presentations and management},
  author={Devine, Michelle F and Kothapalli, Naga and Elkhooly, Mahmoud and Dubey, Divyanshu},
  journal={Therapeutic Advances in Neurological Disorders},
  volume={14},
  pages={1756286420985323},
  year={2021},
  publisher={SAGE Publications Sage UK: London, England}
}

@article{amin2010cardiac,
  title={Cardiac ion channels in health and disease},
  author={Amin, Ahmad S and Tan, Hanno L and Wilde, Arthur AM},
  journal={Heart rhythm},
  volume={7},
  number={1},
  pages={117--126},
  year={2010},
  publisher={Elsevier}
}

@article{cao2025glycolytic,
  title={Glycolytic Dysfunction in Granulosa Cells and Its Contribution to Metabolic Dysfunction in Polycystic Ovary Syndrome},
  author={Cao, Zhenzhen and Zhou, Qin and An, Jie and Guo, Xiaojing and Jia, XiaoFang and Qiu, Yuena},
  journal={Drug Design, Development and Therapy},
  pages={5255--5270},
  year={2025},
  publisher={Taylor \& Francis}
}

@article{d2017electrocardiographic,
  title={Electrocardiographic changes induced by endurance training and pubertal development in male children},
  author={D'Ascenzi, Flavio and Solari, Marco and Anselmi, Francesca and Valentini, Francesca and Barbati, Riccardo and Palmitesta, Paola and Focardi, Marta and Bonifazi, Marco and Mondillo, Sergio},
  journal={The American journal of cardiology},
  volume={119},
  number={5},
  pages={795--801},
  year={2017},
  publisher={Elsevier}
}

@article{stringer2023pathophysiology,
  title={Pathophysiology of ion channels in amyotrophic lateral sclerosis},
  author={Stringer, Robin N and Weiss, Norbert},
  journal={Molecular Brain},
  volume={16},
  number={1},
  pages={82},
  year={2023},
  publisher={Springer}
}

@article{werlein2023inflammation,
  title={Inflammation and vascular remodeling in COVID-19 hearts},
  author={Werlein, Christopher and Ackermann, Maximilian and Stark, Helge and Shah, Harshit R and Tzankov, Alexandar and Haslbauer, Jasmin Dinonne and von Stillfried, Saskia and B{\"u}low, Roman David and El-Armouche, Ali and Kuenzel, Stephan and others},
  journal={Angiogenesis},
  volume={26},
  number={2},
  pages={233--248},
  year={2023},
  publisher={Springer}
}

@inproceedings{iorio2019paraneoplastic,
  title={Paraneoplastic neurological syndromes},
  author={Iorio, Raffaele and Spagni, Gregorio and Masi, Gianvito},
  booktitle={Seminars in diagnostic pathology},
  volume={36},
  number={4},
  pages={279--292},
  year={2019},
  organization={Elsevier}
}

@article{soomro2020paraneoplastic,
  title={Paraneoplastic syndromes in small cell lung cancer},
  author={Soomro, Zaid and Youssef, Michael and Yust-Katz, Shlomit and Jalali, Ali and Patel, Akash J and Mandel, Jacob},
  journal={Journal of Thoracic Disease},
  volume={12},
  number={10},
  pages={6253},
  year={2020}
}

@article{coumbe2018cardiovascular,
  title={Cardiovascular autonomic dysfunction in patients with cancer},
  author={Coumbe, Ben GT and Groarke, John D},
  journal={Current Cardiology Reports},
  volume={20},
  number={8},
  pages={69},
  year={2018},
  publisher={Springer}
}

@article{kamioka2024hypertension,
  title={Hypertension and atrial fibrillation: the clinical impact of hypertension on perioperative outcomes of atrial fibrillation ablation and its optimal control for the prevention of recurrence},
  author={Kamioka, Masashi and Narita, Keisuke and Watanabe, Tomonori and Watanabe, Hiroaki and Makimoto, Hisaki and Okuyama, Takafumi and Yokota, Ayako and Komori, Takahiro and Kabutoya, Tomoyuki and Imai, Yasushi and others},
  journal={Hypertension Research},
  volume={47},
  number={10},
  pages={2800--2810},
  year={2024},
  publisher={Springer Nature Singapore Singapore}
}

@article{rohla2025summary,
  title={Summary of the 2024 ESC-Guidelines for the management of elevated blood pressure and hypertension},
  author={Rohla, Miklos and Rexhaj, Emrush},
  journal={Praxis},
  volume={114},
  number={2},
  pages={48--50},
  year={2025}
}

@article{potpara20242024,
  title={The 2024 European Society of Cardiology guidelines for diagnosis and management of atrial fibrillation: a viewpoint from a practicing clinician's perspective},
  author={Potpara, Tatjana and Romiti, Giulio F and Sohns, Christian},
  journal={Thrombosis and haemostasis},
  volume={124},
  number={12},
  pages={1087--1094},
  year={2024},
  publisher={Georg Thieme Verlag KG}
}

@article{suarez2025hypertension,
  title={Hypertension management in patients with advanced chronic kidney disease with and without dialysis},
  author={Suarez, Maria L Gonzalez and Arriola-Montenegro, Jose and Rol{\'o}n, Leticia},
  journal={Current Opinion in Cardiology},
  volume={40},
  number={4},
  pages={199--205},
  year={2025},
  publisher={LWW}
}

@article{halimi2024management,
  title={Management of patients with hypertension and chronic kidney disease referred to Hypertension Excellence Centres among 27 countries. On behalf of the European Society of Hypertension Working Group on Hypertension and the Kidney},
  author={Halimi, Jean-Michel and Sarafidis, Pantelis and Azizi, Michel and Bilo, Grzegorz and Burkard, Thilo and Bursztyn, Michael and Camafort, Miguel and Chapman, Neil and Cottone, Santina and de Backer, Tine and others},
  journal={Blood pressure},
  volume={33},
  number={1},
  pages={2368800},
  year={2024},
  publisher={Taylor \& Francis}
}

@article{mullen2025recent,
  title={Recent Advances in the Diagnosis and Management of Pulmonary Arterial Hypertension},
  author={Mullen, Eamon and McCullagh, Brian and Gaine, Sean and Quadery, Syed Rehan},
  journal={British Journal of Hospital Medicine},
  volume={86},
  number={2},
  pages={1--13},
  year={2025},
  publisher={MA Healthcare London}
}

@article{rwebembera2024acute,
  title={Acute rheumatic fever and rheumatic heart disease: updates in diagnosis and treatment},
  author={Rwebembera, Joselyn and Beaton, Andrea},
  journal={Current opinion in pediatrics},
  volume={36},
  number={5},
  pages={496--502},
  year={2024},
  publisher={LWW}
}

@article{leitner2025update,
  title={Update COPD and cardiovascular events},
  author={Leitner, Maximilian and Blum, Anna Maria and Bals, Robert},
  journal={Deutsche medizinische Wochenschrift (1946)},
  volume={150},
  number={6},
  pages={298--302},
  year={2025}
}

@article{stieglitz2025copd,
  title={COPD in elderly patients},
  author={Stieglitz, Sven and Frohnhofen, Helmut},
  journal={Pneumologie (Stuttgart, Germany)},
  volume={79},
  number={5},
  pages={382--394},
  year={2025}
}

@article{stracina2022golden,
  title={Golden standard or obsolete method? Review of ECG applications in clinical and experimental context},
  author={Stracina, Tibor and Ronzhina, Marina and Redina, Richard and Novakova, Marie},
  journal={Frontiers in Physiology},
  volume={13},
  pages={867033},
  year={2022},
  publisher={Frontiers Media SA}
}

@article{lippi2021global,
  title={Global epidemiology of atrial fibrillation: an increasing epidemic and public health challenge},
  author={Lippi, Giuseppe and Sanchis-Gomar, Fabian and Cervellin, Gianfranco},
  journal={International journal of stroke},
  volume={16},
  number={2},
  pages={217--221},
  year={2021},
  publisher={SAGE Publications Sage UK: London, England}
}

@article{wang2023early,
  title={Early detection of myocardial ischemia in resting ECG: analysis by HHT},
  author={Wang, Chun-Lin and Wei, Chiu-Chi and Tsai, Cheng-Ting and Lee, Ying-Hsiang and Liu, Lawrence Yu-Min and Chen, Kang-Ying and Lin, Yu-Jen and Lin, Po-Lin},
  journal={BioMedical Engineering OnLine},
  volume={22},
  number={1},
  pages={23},
  year={2023},
  publisher={Springer}
}

@article{kim2025electrocardiography,
  title={Electrocardiography-based artificial intelligence predicts the upcoming future of heart failure with mildly reduced ejection fraction},
  author={Kim, Dae-Young and Lee, Sang-Won and Lee, Dong-Ho and Lee, Sang-Chul and Jang, Ji-Hun and Shin, Sung-Hee and Kim, Dae-Hyeok and Choi, Wonik and Baek, Yong-Soo},
  journal={Frontiers in Cardiovascular Medicine},
  volume={12},
  pages={1418914},
  year={2025},
  publisher={Frontiers Media SA}
}

@misc{attia2023explainable,
  title={Explainable AI for ECG-based prediction of cardiac resynchronization therapy outcomes: learning from machine learning?},
  author={Attia, Zachi I and Friedman, Paul A},
  journal={European heart journal},
  volume={44},
  number={8},
  pages={693--695},
  year={2023},
  publisher={Oxford University Press US}
}

@article{turgut2025unlocking,
  title={Unlocking the diagnostic potential of electrocardiograms through information transfer from cardiac magnetic resonance imaging},
  author={Turgut, {\"O}zg{\"u}n and M{\"u}ller, Philip and Hager, Paul and Shit, Suprosanna and Starck, Sophie and Menten, Martin J and Martens, Eimo and Rueckert, Daniel},
  journal={Medical Image Analysis},
  volume={101},
  pages={103451},
  year={2025},
  publisher={Elsevier}
}

@article{deng2020association,
  title={Association of QTc interval with risk of cardiovascular diseases and related vascular traits: a prospective and longitudinal analysis},
  author={Deng, Chanjuan and Niu, Jingya and Xuan, Liping and Zhu, Wen and Dai, Huajie and Zhao, Zhiyun and Li, Mian and Lu, Jieli and Xu, Yu and Chen, Yuhong and others},
  journal={Global heart},
  volume={15},
  number={1},
  pages={13},
  year={2020}
}

@article{buszman1995use,
  title={Use of changes in ST segment elevation for prediction of infarct artery recanalization in acute myocardial infarction},
  author={Buszman, P and Szafranek, A and Kalarus, Z and Gasior, M},
  journal={European heart journal},
  volume={16},
  number={9},
  pages={1207--1214},
  year={1995},
  publisher={Oxford University Press}
}

@article{chen2022review,
  title={Review of ECG detection and classification based on deep learning: Coherent taxonomy, motivation, open challenges and recommendations},
  author={Chen, Shan Wei and Wang, Shir Li and Qi, Xiu Zhi and Samuri, Suzani Mohamad and Yang, Can},
  journal={Biomedical Signal Processing and Control},
  volume={74},
  pages={103493},
  year={2022},
  publisher={Elsevier}
}

@article{attia2021external,
  title={External validation of a deep learning electrocardiogram algorithm to detect ventricular dysfunction},
  author={Attia, Itzhak Zachi and Tseng, Andrew S and Benavente, Ernest Diez and Medina-Inojosa, Jose R and Clark, Taane G and Malyutina, Sofia and Kapa, Suraj and Schirmer, Henrik and Kudryavtsev, Alexander V and Noseworthy, Peter A and others},
  journal={International journal of cardiology},
  volume={329},
  pages={130--135},
  year={2021},
  publisher={Elsevier}
}

@article{dhingra2025artificial,
  title={Artificial Intelligence--Enabled Prediction of Heart Failure Risk From Single-Lead Electrocardiograms},
  author={Dhingra, Lovedeep S and Aminorroaya, Arya and Pedroso, Aline F and Khunte, Akshay and Sangha, Veer and McIntyre, Daniel and Chow, Clara K and Asselbergs, Folkert W and Brant, Luisa CC and Barreto, Sandhi M and others},
  journal={JAMA cardiology},
  volume={10},
  number={6},
  pages={574--584},
  year={2025},
  publisher={American Medical Association}
}

@article{holmstrom2023deep,
  title={Deep learning-based electrocardiographic screening for chronic kidney disease},
  author={Holmstrom, Lauri and Christensen, Matthew and Yuan, Neal and Weston Hughes, J and Theurer, John and Jujjavarapu, Melvin and Fatehi, Pedram and Kwan, Alan and Sandhu, Roopinder K and Ebinger, Joseph and others},
  journal={Communications Medicine},
  volume={3},
  number={1},
  pages={73},
  year={2023},
  publisher={Nature Publishing Group UK London}
}

@article{kim2024deep,
  title={Deep learning-based long-term risk evaluation of incident type 2 diabetes using electrocardiogram in a non-diabetic population: a retrospective, multicentre study},
  author={Kim, Junmo and Yang, Hyun-Lim and Kim, Su Hwan and Kim, Siun and Lee, Jisoo and Ryu, Jiwon and Kim, Kwangsoo and Kim, Zio and Ahn, Gun and Kwon, Doyun and others},
  journal={Eclinicalmedicine},
  volume={68},
  year={2024},
  publisher={Elsevier}
}

@article{chung2022clinical,
  title={Clinical significance, challenges and limitations in using artificial intelligence for electrocardiography-based diagnosis},
  author={Chung, Cheuk To and Lee, Sharen and King, Emma and Liu, Tong and Armoundas, Antonis A and Bazoukis, George and Tse, Gary},
  journal={International journal of arrhythmia},
  volume={23},
  number={1},
  pages={24},
  year={2022},
  publisher={Springer}
}

@article{siontis2021artificial,
  title={Artificial intelligence-enhanced electrocardiography in cardiovascular disease management},
  author={Siontis, Konstantinos C and Noseworthy, Peter A and Attia, Zachi I and Friedman, Paul A},
  journal={Nature Reviews Cardiology},
  volume={18},
  number={7},
  pages={465--478},
  year={2021},
  publisher={Nature Publishing Group UK London}
}

@article{moreno2024ecg,
  title={ECG-based data-driven solutions for diagnosis and prognosis of cardiovascular diseases: A systematic review},
  author={Moreno-S{\'a}nchez, Pedro A and Garc{\'\i}a-Isla, Guadalupe and Corino, Valentina DA and Vehkaoja, Antti and Brukamp, Kirsten and Van Gils, Mark and Mainardi, Luca},
  journal={Computers in Biology and Medicine},
  volume={172},
  pages={108235},
  year={2024},
  publisher={Elsevier}
}

@article{moosavi2024prospective,
  title={Prospective human validation of artificial intelligence interventions in cardiology: a scoping review},
  author={Moosavi, Amirhossein and Huang, Steven and Vahabi, Maryam and Motamedivafa, Bahar and Tian, Nelly and Mahmood, Rafid and Liu, Peter and Sun, Christopher LF},
  journal={JACC: Advances},
  volume={3},
  number={9\_Part\_2},
  pages={101202},
  year={2024},
  publisher={American College of Cardiology Foundation Washington DC}
}

@article{karabayir2025generalizability,
  title={Generalizability of electrocardiographic artificial intelligence},
  author={Karabayir, Ibrahim and Akbilgic, Oguz},
  journal={npj Cardiovascular Health},
  volume={2},
  number={1},
  pages={38},
  year={2025},
  publisher={Nature Publishing Group UK London}
}

@article{muzammil2024artificial,
  title={Artificial intelligence-enhanced electrocardiography for accurate diagnosis and management of cardiovascular diseases},
  author={Muzammil, Muhammad Ali and Javid, Saman and Afridi, Azra Khan and Siddineni, Rupini and Shahabi, Mariam and Haseeb, Muhammad and Fariha, FNU and Kumar, Satesh and Zaveri, Sahil and Nashwan, Abdulqadir J},
  journal={Journal of electrocardiology},
  volume={83},
  pages={30--40},
  year={2024},
  publisher={Elsevier}
}

@article{moor2023foundation,
  title={Foundation models for generalist medical artificial intelligence},
  author={Moor, Michael and Banerjee, Oishi and Abad, Zahra Shakeri Hossein and Krumholz, Harlan M and Leskovec, Jure and Topol, Eric J and Rajpurkar, Pranav},
  journal={Nature},
  volume={616},
  number={7956},
  pages={259--265},
  year={2023},
  publisher={Nature Publishing Group UK London}
}

@article{ribeiro2021code,
  title={CODE-15\%: A large scale annotated dataset of 12-lead ECGs},
  author={Ribeiro, Ant{\^o}nio H and Paixao, GM and Lima, Emilly M and Ribeiro, M Horta and Pinto Filho, Marcelo M and Gomes, Paulo R and Oliveira, Derick M and Meira Jr, Wagner and Schon, Th{\"o}mas B and Ribeiro, Antonio Luiz P},
  journal={Zenodo, Jun},
  volume={9},
  pages={10--5281},
  year={2021}
}

@article{liu2025biomedical,
  title={Biomedical foundation model: A survey},
  author={Liu, Xiangrui and Zhang, Yuanyuan and Lu, Yingzhou and Yin, Changchang and Hu, Xiaoling and Liu, Xiaoou and Chen, Lulu and Wang, Sheng and Rodriguez, Alexander and Yao, Huaxiu and others},
  journal={arXiv preprint arXiv:2503.02104},
  year={2025}
}

@article{mckeen2024ecg,
  title={Ecg-fm: An open electrocardiogram foundation model},
  author={McKeen, Kaden and Masood, Sameer and Toma, Augustin and Rubin, Barry and Wang, Bo},
  journal={arXiv preprint arXiv:2408.05178},
  year={2024}
}

@article{song2024foundation,
  title={Foundation models for ecg: Leveraging hybrid self-supervised learning for advanced cardiac diagnostics},
  author={Song, Junho and Jang, Jong-Hwan and Lee, Byeong Tak and Hong, DongGyun and Kwon, Joon-myoung and Jo, Yong-Yeon},
  journal={arXiv preprint arXiv:2407.07110},
  year={2024}
}

@article{li2025electrocardiogram,
  title={An Electrocardiogram Foundation Model Built on over 10 Million Recordings},
  author={Li, Jun and Aguirre, Aaron D and Junior, Valdery Moura and Jin, Jiarui and Liu, Che and Zhong, Lanhai and Sun, Chenxi and Clifford, Gari and Brandon Westover, M and Hong, Shenda},
  journal={NEJM AI},
  volume={2},
  number={7},
  pages={AIoa2401033},
  year={2025},
  publisher={Massachusetts Medical Society}
}

@article{koscova2024harvard,
  title={The Harvard-Emory ECG Database},
  author={Koscova, Zuzana and Li, Qiao and Robichaux, Chad and Moura Junior, Valdery and Ghanta, Manohar and Gupta, Aditya and Rosand, Jonathan and Aguirre, Aaron and Hong, Shenda and Albert, David E and others},
  journal={medRxiv},
  pages={2024--09},
  year={2024},
  publisher={Cold Spring Harbor Laboratory Press}
}

@article{ptbxl,
  title={PTB-XL, a large publicly available electrocardiography dataset},
  author={Wagner, Patrick and Strodthoff, Nils and Bousseljot, Ralf-Dieter and Kreiseler, Dieter and Lunze, Fatima I and Samek, Wojciech and Schaeffter, Tobias},
  journal={Scientific data},
  volume={7},
  number={1},
  pages={1--15},
  year={2020},
  publisher={Nature Publishing Group}
}

@misc{ukbb,
  title={UK biobank data: come and get it},
  author={Allen, Naomi E and Sudlow, Cathie and Peakman, Tim and Collins, Rory and Uk biobank},
  journal={Science translational medicine},
  volume={6},
  number={224},
  pages={224ed4--224ed4},
  year={2014},
  publisher={American Association for the Advancement of Science}
}

\bigskip


\newpage

\appendix
\setcounter{table}{0} 
\renewcommand{\thetable}{A\arabic{table}}

\section{Diagnostic performance of AnyECG for ICD-10 diagnoses}
The diagnostic performance of AnyECG for ICD-10 codes with an AUC of at least 0.65 is summarized in Table~\ref{tab:icd_data}, which covers ICD chapters I to XV and includes a total of 588 three-digit ICD-10 codes.

\begin{longtable}{p{1cm} p{8.8cm} r r r}
\caption{Summary of ICD-10 codes with an AUC greater than 0.65 for AnyECG. ICD, International Classification of Diseases; AUC, area under the receiver operating characteristic curve.}
\label{tab:icd_data} \\
\toprule
\textbf{ICD} & \textbf{Description} & \textbf{No. Positive} & \textbf{No. Negative} & \textbf{AUC} \\
\midrule
\endfirsthead

\multicolumn{5}{c}{\tablename\ \thetable{} -- Continued from previous page} \\
\toprule
\textbf{ICD} & \textbf{Description} & \textbf{No. Positive} & \textbf{No. Negative} & \textbf{AUC} \\
\midrule
\endhead

\midrule
\multicolumn{5}{r}{Continued on next page} \\
\bottomrule
\endfoot

\bottomrule
\endlastfoot
\rowcolor{gray!20} 
\multicolumn{5}{l}{I. Certain infectious and parasitic diseases} \\
A04  & Other bacterial intestinal infections   & 4538& 203478   & 0.69  \\
A07  & Other protozoal intestinal diseases& 77  & 207939   & 0.86  \\
A17  & Tuberculosis of nervous system& 16  & 208000   & 0.66  \\
A18  & Tuberculosis of other organs  & 46  & 207970   & 0.69  \\
A28  & Other zoonotic bacterial diseases, not elsewhere classified & 20  & 207996   & 0.80  \\
A31  & Infection due to other mycobacteria& 622 & 207394   & 0.66  \\
A37  & Whooping cough & 6   & 208010   & 0.70  \\
A38  & Scarlet fever  & 5   & 208011   & 0.75  \\
A41  & Other sepsis   & 9615& 198401   & 0.72  \\
A43  & Nocardiosis    & 181 & 207835   & 0.69  \\
A46  & Erysipelas& 57  & 207959   & 0.70  \\
A49  & Bacterial infection of unspecified site & 4226& 203790   & 0.66  \\
A51  & Early syphilis & 39  & 207977   & 0.84  \\
A56  & Other sexually transmitted chlamydial diseases    & 10  & 208006   & 0.75  \\
A58  & Granuloma inguinale & 1   & 208015   & 0.75  \\
A59  & Trichomoniasis & 49  & 207967   & 0.82  \\
A60  & Anogenital herpesviral {[}herpes simplex{]} infections & 638 & 207378   & 0.67  \\
A64  & Unspecified sexually transmitted disease& 149 & 207867   & 0.68  \\
A68  & Relapsing fevers    & 41  & 207975   & 0.85  \\
A74  & Other diseases caused by chlamydiae& 181 & 207835   & 0.84  \\
A78  & Q fever   & 1   & 208015   & 0.75  \\
A81  & Atypical virus infections of central nervous system    & 10  & 208006   & 0.78  \\
A87  & Viral meningitis    & 20  & 207996   & 0.72  \\
A93  & Other arthropod-borne viral fevers, not elsewhere classified& 66  & 207950   & 0.69  \\
B10  & Other human herpesviruses& 6   & 208010   & 0.66  \\
B15  & Acute hepatitis A   & 10  & 208006   & 0.68  \\
B17  & Other acute viral hepatitis   & 476 & 207540   & 0.69  \\
B18  & Chronic viral hepatitis  & 2877& 205139   & 0.66  \\
B25  & Cytomegaloviral disease  & 1441& 206575   & 0.70  \\
B27  & Infectious mononucleosis & 242 & 207774   & 0.74  \\
B34  & Viral infection of unspecified site& 4433& 203583   & 0.66  \\
B39  & Histoplasmosis & 17  & 207999   & 0.66  \\
B44  & Aspergillosis  & 1259& 206757   & 0.72  \\
B45  & Cryptococcosis & 97  & 207919   & 0.79  \\
B46  & Zygomycosis    & 42  & 207974   & 0.92  \\
B49  & Unspecified mycosis & 2197& 205819   & 0.71  \\
B58  & Toxoplasmosis  & 148 & 207868   & 0.72  \\
B59  & Pneumocystosis & 277 & 207739   & 0.72  \\
B71  & Other cestode infections & 4   & 208012   & 0.90  \\
B89  & Unspecified parasitic disease & 3146& 204870   & 0.66  \\
B95  & Streptococcus, Staphylococcus, and Enterococcus as the cause of diseases classified elsewhere  & 4023& 203993   & 0.66  \\
B99  & Other and unspecified infectious diseases    & 3264& 204752   & 0.66  \\

\midrule
\rowcolor{gray!20} 
\multicolumn{5}{l}{II. Neoplasms} \\
C00  & Malignant neoplasm of lip& 15  & 208001   & 0.67    \\
C01  & Malignant neoplasm of base of tongue    & 336 & 207680   & 0.70    \\
C07  & Malignant neoplasm of parotid gland& 115 & 207901   & 0.71    \\
C10  & Malignant neoplasm of oropharynx   & 762 & 207254   & 0.67    \\
C12  & Malignant neoplasm of pyriform sinus    & 59  & 207957   & 0.82    \\
C13  & Malignant neoplasm of hypopharynx  & 133 & 207883   & 0.76    \\
C15  & Malignant neoplasm of esophagus    & 1294& 206722   & 0.86    \\
C16  & Malignant neoplasm of stomach & 903 & 207113   & 0.74    \\
C19  & Malignant neoplasm of rectosigmoid junction  & 606 & 207410   & 0.76    \\
C20  & Malignant neoplasm of rectum  & 1300& 206716   & 0.66    \\
C22  & Malignant neoplasm of liver and intrahepatic bile ducts& 1543& 206473   & 0.69    \\
C24  & Malignant neoplasm of other and unspecified parts of biliary tract    & 241 & 207775   & 0.75    \\
C25  & Malignant neoplasm of pancreas& 1808& 206208   & 0.67    \\
C26  & Malignant neoplasm of other and ill-defined digestive organs& 36  & 207980   & 0.75    \\
C30  & Malignant neoplasm of nasal cavity and middle ear & 91  & 207925   & 0.70    \\
C31  & Malignant neoplasm of accessory sinuses & 111 & 207905   & 0.67    \\
C33  & Malignant neoplasm of trachea & 56  & 207960   & 0.75    \\
C34  & Malignant neoplasm of bronchus and lung & 7297& 200719   & 0.65    \\
C37  & Malignant neoplasm of thymus  & 72  & 207944   & 0.67    \\
C38  & Malignant neoplasm of heart, mediastinum and pleura    & 438 & 207578   & 0.70    \\
C40  & Malignant neoplasm of bone and articular cartilage of limbs & 151 & 207865   & 0.76    \\
C41  & Malignant neoplasm of bone and articular cartilage of other and unspecified sites    & 683 & 207333   & 0.84    \\
C43  & Malignant melanoma of skin    & 2715& 205301   & 0.68    \\
C44  & Other and unspecified malignant neoplasm of skin  & 8754& 199262   & 0.72    \\
C45  & Mesothelioma   & 471 & 207545   & 0.66    \\
C47  & Malignant neoplasm of peripheral nerves and autonomic nervous system  & 55  & 207961   & 0.77    \\
C48  & Malignant neoplasm of retroperitoneum and peritoneum   & 458 & 207558   & 0.71    \\
C49  & Malignant neoplasm of other connective and soft tissue & 2172& 205844   & 0.66    \\
C4A  & Merkel cell carcinoma    & 110 & 207906   & 0.70    \\
C50  & Malignant neoplasm of breast  & 6721& 201295   & 0.71    \\
C52  & Malignant neoplasm of vagina  & 79  & 207937   & 0.78    \\
C53  & Malignant neoplasm of cervix uteri & 374 & 207642   & 0.67    \\
C54  & Malignant neoplasm of corpus uteri & 1394& 206622   & 0.70    \\
C56  & Malignant neoplasm of ovary   & 1383& 206633   & 0.81    \\
C57  & Malignant neoplasm of other and unspecified female genital organs& 283 & 207733   & 0.72    \\
C60  & Malignant neoplasm of penis   & 37  & 207979   & 0.76    \\
C61  & Malignant neoplasm of prostate& 7370& 200646   & 0.74    \\
C65  & Malignant neoplasm of renal pelvis & 234 & 207782   & 0.78    \\
C66  & Malignant neoplasm of ureter  & 300 & 207716   & 0.75    \\
C67  & Malignant neoplasm of bladder & 2917& 205099   & 0.68    \\
C70  & Malignant neoplasm of meninges& 8   & 208008   & 0.73    \\
C71  & Malignant neoplasm of brain   & 1412& 206604   & 0.68    \\
C78  & Secondary malignant neoplasm of respiratory and digestive organs & 3679& 204337   & 0.66    \\
C81  & Hodgkin lymphoma    & 1152& 206864   & 0.66    \\
C83  & Non-follicular lymphoma  & 3445& 204571   & 0.66    \\
C85  & Other specified and unspecified types of non-Hodgkin lymphoma    & 4594& 203422   & 0.66    \\
C88  & Malignant immunoproliferative diseases and certain other B-cell lymphomas  & 640 & 207376   & 0.66    \\
C90  & Multiple myeloma and malignant plasma cell neoplasms   & 2495& 205521   & 0.66    \\
C91  & Lymphoid leukemia   & 2874& 205142   & 0.85    \\
C92  & Myeloid leukemia    & 3581& 204435   & 0.71    \\
C94  & Other leukemias of specified cell type  & 340 & 207676   & 0.77    \\
C95  & Leukemia of unspecified cell type  & 1399& 206617   & 0.79    \\
D03  & Melanoma in situ    & 623 & 207393   & 0.71    \\
D04  & Carcinoma in situ of skin& 1390& 206626   & 0.74    \\
D05  & Carcinoma in situ of breast   & 1071& 206945   & 0.71    \\
D06  & Carcinoma in situ of cervix uteri  & 71  & 207945   & 0.75    \\
D09  & Carcinoma in situ of other and unspecified sites  & 519 & 207497   & 0.72    \\
D11  & Benign neoplasm of major salivary glands& 131 & 207885   & 0.67    \\
D18  & Hemangioma and lymphangioma, any site   & 9065& 198951   & 0.67    \\
D21  & Other benign neoplasms of connective and other soft tissue  & 843 & 207173   & 0.71    \\
D22  & Melanocytic nevi    & 18638    & 189378   & 0.65    \\
D24  & Benign neoplasm of breast& 151 & 207865   & 0.68    \\
D25  & Leiomyoma of uterus & 1037& 206979   & 0.76    \\
D29  & Benign neoplasm of male genital organs  & 36  & 207980   & 0.70    \\
D31  & Benign neoplasm of eye and adnexa  & 1660& 206356   & 0.68    \\
D38  & Neoplasm of uncertain behavior of middle ear and respiratory and intrathoracic organs& 145 & 207871   & 0.71    \\
D39  & Neoplasm of uncertain behavior of female genital organs& 68  & 207948   & 0.76    \\
D41  & Neoplasm of uncertain behavior of urinary organs  & 118 & 207898   & 0.70    \\
D43  & Neoplasm of uncertain behavior of brain and central nervous system    & 74  & 207942   & 0.65    \\
D46  & Myelodysplastic syndromes& 2172& 205844   & 0.69    \\
D48  & Neoplasm of uncertain behavior of other and unspecified sites    & 11434    & 196582   & 0.69    \\

\midrule
\rowcolor{gray!20} 
\multicolumn{5}{l}{\makecell[l]{III. Diseases of the blood and blood-forming organs and certain disorders involving the immune \\ mechanism}} \\
D50  & Iron deficiency anemia   & 22100    & 185916   & 0.83    \\
D51  & Vitamin B12 deficiency anemia & 951 & 207065   & 0.69    \\
D52  & Folate deficiency anemia & 127 & 207889   & 0.73    \\
D53  & Other nutritional anemias& 1751& 206265   & 0.66    \\
D55  & Anemia due to enzyme disorders& 19  & 207997   & 0.77    \\
D57  & Sickle-cell disorders    & 764 & 207252   & 0.78    \\
D58  & Other hereditary hemolytic anemias & 517 & 207499   & 0.67    \\
D59  & Acquired hemolytic anemia& 637 & 207379   & 0.70    \\
D61  & Other aplastic anemias and other bone marrow failure syndromes   & 4953& 203063   & 0.71    \\
D62  & Acute posthemorrhagic anemia  & 1870& 206146   & 0.65    \\
D63  & Anemia in chronic diseases classified elsewhere   & 9162& 198854   & 0.72    \\
D64  & Other anemias  & 37738    & 170278   & 0.74    \\
D65  & Disseminated intravascular coagulation {[}defibrination syndrome{]}   & 98  & 207918   & 0.73    \\
D66  & Hereditary factor VIII deficiency  & 48  & 207968   & 0.83    \\
D70  & Neutropenia    & 5568& 202448   & 0.70    \\
D71  & Functional disorders of polymorphonuclear neutrophils  & 87  & 207929   & 0.68    \\
D73  & Diseases of spleen  & 1120& 206896   & 0.73    \\
D76  & Other specified diseases with participation of lymphoreticular and reticulohistiocytic tissue  & 56  & 207960   & 0.89    \\
D80  & Immunodeficiency with predominantly antibody defects   & 1965& 206051   & 0.67    \\
D84  & Other immunodeficiencies & 8544& 199472   & 0.67    \\

\midrule
\rowcolor{gray!20} 
\multicolumn{5}{l}{IV. Endocrine, nutritional and metabolic diseases} \\
E02  & Subclinical iodine-deficiency hypothyroidism & 81  & 207935   & 0.71  \\
E08  & Diabetes mellitus due to underlying condition& 6980& 201036   & 0.70  \\
E09  & Drug or chemical induced diabetes mellitus   & 2559& 205457   & 0.70  \\
E10  & Type 1 diabetes mellitus & 2752& 205264   & 0.69  \\
E11  & Type 2 diabetes mellitus & 39367    & 168649   & 0.70  \\
E13  & Other specified diabetes mellitus  & 10846    & 197170   & 0.71  \\
E16  & Other disorders of pancreatic internal secretion  & 4051& 203965   & 0.68  \\
E21  & Hyperparathyroidism and other disorders of parathyroid gland& 5167& 202849   & 0.94  \\
E24  & Cushing's syndrome  & 294 & 207722   & 0.72  \\
E26  & Hyperaldosteronism  & 325 & 207691   & 0.69  \\
E28  & Ovarian dysfunction & 1727& 206289   & 0.78  \\
E29  & Testicular dysfunction   & 1529& 206487   & 0.72  \\
E30  & Disorders of puberty, not elsewhere classified    & 23  & 207993   & 0.87  \\
E32  & Diseases of thymus  & 68  & 207948   & 0.75  \\
E40  & Kwashiorkor    & 218 & 207798   & 0.65  \\
E43  & Unspecified severe protein-calorie malnutrition   & 3645& 204371   & 0.76  \\
E44  & Protein-calorie malnutrition of moderate and mild degree    & 2449& 205567   & 0.70  \\
E46  & Unspecified protein-calorie malnutrition& 4893& 203123   & 0.71  \\
E50  & Vitamin A deficiency& 278 & 207738   & 0.69  \\
E51  & Thiamine deficiency & 194 & 207822   & 0.70  \\
E53  & Deficiency of other B group vitamins    & 6236& 201780   & 0.81  \\
E56  & Other vitamin deficiencies    & 1935& 206081   & 0.70  \\
E59  & Dietary selenium deficiency   & 16  & 208000   & 0.69  \\
E60  & Dietary zinc deficiency  & 343 & 207673   & 0.66  \\
E63  & Other nutritional deficiencies& 1415& 206601   & 0.71  \\
E64  & Sequelae of malnutrition and other nutritional deficiencies & 32  & 207984   & 0.73  \\
E65  & Localized adiposity & 321 & 207695   & 0.67  \\
E66  & Overweight and obesity   & 31279    & 176737   & 0.71  \\
E72  & Other disorders of amino-acid metabolism& 290 & 207726   & 0.77  \\
E76  & Disorders of glycosaminoglycan metabolism    & 6   & 208010   & 0.89  \\
E78  & Disorders of lipoprotein metabolism and other lipidemias    & 80058    & 127958   & 0.71  \\
E79  & Disorders of purine and pyrimidine metabolism& 1032& 206984   & 0.75  \\
E83  & Disorders of mineral metabolism    & 16980    & 191036   & 0.88  \\
E84  & Cystic fibrosis& 370 & 207646   & 0.77  \\
E85  & Amyloidosis    & 4061& 203955   & 0.76  \\
E87  & Other disorders of fluid, electrolyte and acid-base balance & 41520    & 166496   & 0.71  \\
E95  & & 677 & 207339   & 0.80  \\
E96  & & 6   & 208010   & 0.82  \\
\midrule
\rowcolor{gray!20} 
\multicolumn{5}{l}{V. Mental and behavioural disorders} \\
F01  & Vascular dementia   & 830 & 207186   & 0.75  \\
F02  & Dementia in other diseases classified elsewhere   & 2145& 205871   & 0.70  \\
F03  & Unspecified dementia& 2420& 205596   & 0.72  \\
F05  & Delirium due to known physiological condition& 850 & 207166   & 0.73  \\
F10  & Alcohol related disorders& 10995    & 197021   & 0.67  \\
F11  & Opioid related disorders & 7803& 200213   & 0.69  \\
F12  & Cannabis related disorders    & 1256& 206760   & 0.74  \\
F13  & Sedative, hypnotic, or anxiolytic related disorders    & 1121& 206895   & 0.72  \\
F14  & Cocaine related disorders& 1838& 206178   & 0.70  \\
F15  & Other stimulant related disorders  & 791 & 207225   & 0.75  \\
F16  & Hallucinogen related disorders& 61  & 207955   & 0.82  \\
F18  & Inhalant related disorders    & 4903& 203113   & 0.73  \\
F19  & Other psychoactive substance related disorders    & 7052& 200964   & 0.70  \\
F25  & Schizoaffective disorders& 1055& 206961   & 0.68  \\
F28  & Other psychotic disorder not due to a substance or known physiological condition& 50  & 207966   & 0.70  \\
F30  & Manic episode  & 346 & 207670   & 0.69  \\
F31  & Bipolar disorder    & 4005& 204011   & 0.66  \\
F42  & Obsessive-compulsive disorder & 869 & 207147   & 0.66  \\
F44  & Dissociative and conversion disorders   & 825 & 207191   & 0.75  \\
F45  & Somatoform disorders& 699 & 207317   & 0.65  \\
F48  & Other nonpsychotic mental disorders& 1773& 206243   & 0.66  \\
F50  & Eating disorders    & 2453& 205563   & 0.71  \\
F53  & Mental and behavioral disorders associated with the puerperium, not elsewhere classified  & 149 & 207867   & 0.83  \\
F55  & Abuse of non-psychoactive substances    & 17  & 207999   & 0.68  \\
F59  & Unspecified behavioral syndromes associated with physiological disturbances and physical factors    & 4   & 208012   & 0.95  \\
F60  & Specific personality disorders& 1096& 206920   & 0.71  \\
F63  & Impulse disorders   & 2034& 205982   & 0.66  \\
F64  & Gender identity disorders& 129 & 207887   & 0.73  \\
F68  & Other disorders of adult personality and behavior & 71  & 207945   & 0.66  \\
F69  & Unspecified disorder of adult personality and behavior & 79  & 207937   & 0.65  \\
F70  & Mild intellectual disabilities& 25  & 207991   & 0.72  \\
F73  & Profound intellectual disabilities & 3   & 208013   & 0.88  \\
F79  & Unspecified intellectual disabilities   & 437 & 207579   & 0.71  \\
F80  & Specific developmental disorders of speech and language& 313 & 207703   & 0.75  \\
F81  & Specific developmental disorders of scholastic skills  & 292 & 207724   & 0.66  \\
F82  & Specific developmental disorder of motor function & 45  & 207971   & 0.87  \\
F84  & Pervasive developmental disorders  & 383 & 207633   & 0.73  \\
F88  & Other disorders of psychological development & 38  & 207978   & 0.86  \\
F89  & Unspecified disorder of psychological development & 58  & 207958   & 0.70  \\
F90  & Attention-deficit hyperactivity disorders    & 3730& 204286   & 0.70  \\
F93  & Emotional disorders with onset specific to childhood   & 9   & 208007   & 0.82  \\
F94  & Disorders of social functioning with onset specific to childhood and adolescence& 5   & 208011   & 0.79  \\
F95  & Tic disorder   & 80  & 207936   & 0.68  \\
F98  & Other behavioral and emotional disorders with onset usually occurring in childhood and adolescence  & 1063& 206953   & 0.67  \\
F99  & Mental disorder, not otherwise specified& 1694& 206322   & 0.70  \\
\midrule
\rowcolor{gray!20} 
\multicolumn{5}{l}{VI. Diseases of the nervous system} \\
G00  & Bacterial meningitis, not elsewhere classified    & 16  & 208000   & 0.65  \\
G01  & Meningitis in bacterial diseases classified elsewhere  & 23  & 207993   & 0.73  \\
G03  & Meningitis due to other and unspecified causes    & 524 & 207492   & 0.66  \\
G04  & Encephalitis, myelitis and encephalomyelitis & 712 & 207304   & 0.68  \\
G08  & Intracranial and intraspinal phlebitis and thrombophlebitis & 486 & 207530   & 0.76  \\
G09  & Sequelae of inflammatory diseases of central nervous system & 5   & 208011   & 0.67  \\
G10  & Huntington's disease& 14  & 208002   & 0.80  \\
G12  & Spinal muscular atrophy and related syndromes& 273 & 207743   & 0.71  \\
G14  & Postpolio syndrome  & 46  & 207970   & 0.69  \\
G20  & Parkinson's disease & 1874& 206142   & 0.76  \\
G23  & Other degenerative diseases of basal ganglia & 265 & 207751   & 0.68  \\
G30  & Alzheimer's disease & 1393& 206623   & 0.71  \\
G31  & Other degenerative diseases of nervous system, not elsewhere classified    & 2970& 205046   & 0.80  \\
G37  & Other demyelinating diseases of central nervous system & 299 & 207717   & 0.69  \\
G43  & Migraine  & 9325& 198691   & 0.76  \\
G45  & Transient cerebral ischemic attacks and related syndromes   & 5929& 202087   & 0.75  \\
G47  & Sleep disorders& 38056    & 169960   & 0.68  \\
G53  & Cranial nerve disorders in diseases classified elsewhere    & 9   & 208007   & 0.89  \\
G59  & Mononeuropathy in diseases classified elsewhere   & 11  & 208005   & 0.68  \\
G63  & Polyneuropathy in diseases classified elsewhere   & 949 & 207067   & 0.67  \\
G65  & Sequelae of inflammatory and toxic polyneuropathies    & 5   & 208011   & 0.72  \\
G72  & Other and unspecified myopathies   & 1360& 206656   & 0.67  \\
G73  & Disorders of myoneural junction and muscle in diseases classified elsewhere& 16  & 208000   & 0.66  \\
G80  & Cerebral palsy & 188 & 207828   & 0.76  \\
G82  & Paraplegia (paraparesis) and quadriplegia (quadriparesis)   & 1221& 206795   & 0.71  \\
G90  & Disorders of autonomic nervous system   & 2506& 205510   & 0.66  \\
G99  & Other disorders of nervous system in diseases classified elsewhere    & 763 & 207253   & 0.73  \\
\midrule
\rowcolor{gray!20} 
\multicolumn{5}{l}{VII. Diseases of the eye and adnexa} \\

H01  & Other inflammation of eyelid  & 4351& 203665   & 0.67  \\
H16  & Keratitis & 1645& 206371   & 0.68  \\
H17  & Corneal scars and opacities   & 543 & 207473   & 0.68  \\
H21  & Other disorders of iris and ciliary body& 647 & 207369   & 0.66  \\
H25  & Age-related cataract& 20273    & 187743   & 0.69  \\
H26  & Other cataract & 7856& 200160   & 0.70  \\
H27  & Other disorders of lens  & 288 & 207728   & 0.80  \\
H30  & Chorioretinal inflammation    & 153 & 207863   & 0.70  \\
H31  & Other disorders of choroid    & 1296& 206720   & 0.71  \\
H32  & Chorioretinal disorders in diseases classified elsewhere    & 6   & 208010   & 0.90  \\
H34  & Retinal vascular occlusions   & 2193& 205823   & 0.69  \\
H35  & Other retinal disorders  & 12142    & 195874   & 0.68  \\
H36  & Retinal disorders in diseases classified elsewhere& 82  & 207934   & 0.96  \\
H40  & Glaucoma  & 12452    & 195564   & 0.66  \\
H42  & Glaucoma in diseases classified elsewhere    & 20  & 207996   & 0.68  \\
H43  & Disorders of vitreous body    & 8807& 199209   & 0.68  \\
H46  & Optic neuritis & 740 & 207276   & 0.76  \\
H61  & Other disorders of external ear    & 9860& 198156   & 0.67  \\
H65  & Nonsuppurative otitis media   & 1785& 206231   & 0.66  \\
H66  & Suppurative and unspecified otitis media& 1857& 206159   & 0.69  \\
H68  & Eustachian salpingitis and obstruction  & 12  & 208004   & 0.69  \\
H70  & Mastoiditis and related conditions & 157 & 207859   & 0.76  \\
H71  & Cholesteatoma of middle ear   & 191 & 207825   & 0.76  \\
H74  & Other disorders of middle ear mastoid   & 193 & 207823   & 0.69  \\
H80  & Otosclerosis   & 123 & 207893   & 0.68  \\
H81  & Disorders of vestibular function   & 4017& 203999   & 0.78  \\
H90  & Conductive and sensorineural hearing loss    & 13695    & 194321   & 0.68  \\
H91  & Other and unspecified hearing loss & 9658& 198358   & 0.68  \\

\midrule
\rowcolor{gray!20} 
\multicolumn{5}{l}{VIII. Diseases of the ear and mastoid process} \\
H73 & Other disorders of tympanic membrane & 140   & 359760 & 0.72 \\
H90 & Conductive and sensorineural hearing loss & 5640 & 354260 & 0.71 \\
H80 & Otosclerosis& 180   & 359720 & 0.69 \\
H91 & Other and unspecified hearing loss   & 4660 & 355240 & 0.68 \\
H61 & Other disorders of external ear & 2360 & 357540 & 0.66 \\
\midrule
\rowcolor{gray!20} 
\multicolumn{5}{l}{IX. Diseases of the circulatory system} \\
I01  & Rheumatic fever with heart involvement  & 289 & 207727   & 0.71  \\
I05  & Rheumatic mitral valve diseases    & 5545& 202471   & 0.88  \\
I06  & Rheumatic aortic valve diseases    & 959 & 207057   & 0.89  \\
I07  & Rheumatic tricuspid valve diseases & 3259& 204757   & 0.86  \\
I08  & Multiple valve diseases  & 559 & 207457   & 0.93  \\
I09  & Other rheumatic heart diseases& 391 & 207625   & 0.93  \\
I10  & Essential (primary) hypertension   & 97583    & 110433   & 0.82  \\
I11  & Hypertensive heart disease    & 2816& 205200   & 0.82  \\
I12  & Hypertensive chronic kidney disease& 3195& 204821   & 0.88  \\
I13  & Hypertensive heart and chronic kidney disease& 2332& 205684   & 0.76  \\
I15  & Secondary hypertension   & 8638& 199378   & 0.77  \\
I16  & Hypertensive crisis & 2772& 205244   & 0.71  \\
I20  & Angina pectoris& 26244    & 181772   & 0.72  \\
I21  & Acute myocardial infarction   & 20255    & 187761   & 0.95  \\
I22  & Subsequent ST elevation (STEMI) and non-ST elevation (NSTEMI) myocardial infarction  & 191 & 207825   & 0.69  \\
I23  & Certain current complications following ST elevation (STEMI) and non-ST elevation (NSTEMI) myocardial infarction (within the 28 day period) & 288 & 207728   & 0.96  \\
I24  & Other acute ischemic heart diseases& 6601& 201415   & 0.81  \\
I25  & Chronic ischemic heart disease& 54195    & 153821   & 0.76  \\
I27  & Other pulmonary heart diseases& 11242    & 196774   & 0.71  \\
I31  & Other diseases of pericardium & 11668    & 196348   & 0.71  \\
I32  & Pericarditis in diseases classified elsewhere& 132 & 207884   & 0.73  \\
I33  & Acute and subacute endocarditis    & 6462& 201554   & 0.78  \\
I34  & Nonrheumatic mitral valve disorders& 13311    & 194705   & 0.88  \\
I35  & Nonrheumatic aortic valve disorders& 18469    & 189547   & 0.91  \\
I36  & Nonrheumatic tricuspid valve disorders  & 3471& 204545   & 0.91  \\
I37  & Nonrheumatic pulmonary valve disorders  & 812 & 207204   & 0.86  \\
I38  & Endocarditis, valve unspecified    & 5048& 202968   & 0.90  \\
I39  & Endocarditis and heart valve disorders in diseases classified elsewhere    & 167 & 207849   & 0.81  \\
I42  & Cardiomyopathy & 26650    & 181366   & 0.95  \\
I43  & Cardiomyopathy in diseases classified elsewhere   & 1881& 206135   & 0.83  \\
I44  & Atrioventricular and left bundle-branch block& 11488    & 196528   & 0.92  \\
I45  & Other conduction disorders    & 9159& 198857   & 0.91  \\
I46  & Cardiac arrest & 5423& 202593   & 0.77  \\
I47  & Paroxysmal tachycardia & 23608 & 184408   & 0.85  \\
I48  & Atrial fibrillation and flutter    & 55138    & 152878   & 0.95  \\
I49  & Other cardiac arrhythmias& 71655    & 136361   & 0.71  \\
I50  & Heart failure  & 53172    & 154844   & 0.83  \\
I60  & Nontraumatic subarachnoid hemorrhage    & 1725& 206291   & 0.70  \\
I62  & Other and unspecified nontraumatic intracranial hemorrhage  & 5100& 202916   & 0.70  \\
I63  & Cerebral infarction & 17773    & 190243   & 0.71  \\
I65  & Occlusion and stenosis of precerebral arteries, not resulting in cerebral infarction & 8115& 199901   & 0.72  \\
I66  & Occlusion and stenosis of cerebral arteries, not resulting in cerebral infarction    & 315 & 207701   & 0.73  \\
I67  & Other cerebrovascular diseases& 9082& 198934   & 0.73  \\
I68  & Cerebrovascular disorders in diseases classified elsewhere  & 308 & 207708   & 0.70  \\
I69  & Sequelae of cerebrovascular disease& 2454& 205562   & 0.76  \\
I70  & Atherosclerosis& 9515& 198501   & 0.70  \\
I71  & Aortic aneurysm and dissection& 10134    & 197882   & 0.75  \\
I73  & Other peripheral vascular diseases & 14279    & 193737   & 0.72  \\
I74  & Arterial embolism and thrombosis   & 2504& 205512   & 0.75  \\
I75  & Atheroembolism & 81  & 207935   & 0.67  \\
I76  & Septic arterial embolism & 214 & 207802   & 0.75  \\
I78  & Diseases of capillaries  & 1055& 206961   & 0.67  \\
I79  & Disorders of arteries, arterioles and capillaries in diseases classified elsewhere   & 3239& 204777   & 0.77  \\
I81  & Portal vein thrombosis   & 603 & 207413   & 0.77  \\
I82  & Other venous embolism and thrombosis    & 16623    & 191393   & 0.79  \\
I85  & Esophageal varices  & 1375& 206641   & 0.75  \\
I96  & Gangrene, not elsewhere classified & 1818& 206198   & 0.75  \\
I97  & Intraoperative and postprocedural complications and disorders of circulatory system, not elsewhere classified & 1037& 206979   & 0.82  \\

\midrule
\rowcolor{gray!20} 
\multicolumn{5}{l}{X. Diseases of the respiratory system} \\

J00  & Acute nasopharyngitis {[}common cold{]} & 402 & 207614   & 0.76  \\
J01  & Acute sinusitis& 5576& 202440   & 0.65  \\
J03  & Acute tonsillitis   & 293 & 207723   & 0.72  \\
J05  & Acute obstructive laryngitis {[}croup{]} and epiglottitis   & 31  & 207985   & 0.67  \\
J06  & Acute upper respiratory infections of multiple and unspecified sites  & 13780    & 194236   & 0.69  \\
J12  & Viral pneumonia, not elsewhere classified    & 3581& 204435   & 0.66  \\
J14  & Pneumonia due to Hemophilus influenzae  & 150 & 207866   & 0.75  \\
J15  & Bacterial pneumonia, not elsewhere classified& 1983& 206033   & 0.67  \\
J16  & Pneumonia due to other infectious organisms, not elsewhere classified & 569 & 207447   & 0.74  \\
J17  & Pneumonia in diseases classified elsewhere   & 522 & 207494   & 0.77  \\
J21  & Acute bronchiolitis & 642 & 207374   & 0.67  \\
J30  & Vasomotor and allergic rhinitis    & 10397    & 197619   & 0.75  \\
J41  & Simple and mucopurulent chronic bronchitis   & 2702& 205314   & 0.69  \\
J42  & Unspecified chronic bronchitis& 2812& 205204   & 0.69  \\
J43  & Emphysema & 7288& 200728   & 0.72  \\
J44  & Other chronic obstructive pulmonary disease  & 14586    & 193430   & 0.77  \\
J47  & Bronchiectasis & 2564& 205452   & 0.65  \\
J61  & Pneumoconiosis due to asbestos and other mineral fibers& 110 & 207906   & 0.73  \\
J63  & Pneumoconiosis due to other inorganic dusts  & 5   & 208011   & 0.77  \\
J67  & Hypersensitivity pneumonitis due to organic dust  & 201 & 207815   & 0.75  \\
J69  & Pneumonitis due to solids and liquids   & 3929& 204087   & 0.66  \\
J80  & Acute respiratory distress syndrome& 4105& 203911   & 0.74  \\
J81  & Pulmonary edema& 9663& 198353   & 0.76  \\
J82  & Pulmonary eosinophilia, not elsewhere classified  & 103 & 207913   & 0.67  \\
J84  & Other interstitial pulmonary diseases   & 7742& 200274   & 0.67  \\
J85  & Abscess of lung and mediastinum    & 444 & 207572   & 0.76  \\
J86  & Pyothorax & 1127& 206889   & 0.74  \\
J90  & Pleural effusion, not elsewhere classified   & 12465    & 195551   & 0.70  \\
J91  & Pleural effusion in conditions classified elsewhere    & 12341    & 195675   & 0.70  \\
J92  & Pleural plaque & 301 & 207715   & 0.69  \\
J93  & Pneumothorax and air leak& 1276& 206740   & 0.68  \\
J95  & Intraoperative and postprocedural complications and disorders of respiratory system, not elsewhere classified & 2636& 205380   & 0.70  \\
J96  & Respiratory failure, not elsewhere classified& 17510    & 190506   & 0.74  \\

\midrule
\rowcolor{gray!20} 
\multicolumn{5}{l}{XI. Diseases of the digestive system} \\
K01  & Embedded and impacted teeth   & 482 & 207534   & 0.66  \\
K04  & Diseases of pulp and periapical tissues & 2466& 205550   & 0.70  \\
K06  & Other disorders of gingiva and edentulous alveolar ridge    & 372 & 207644   & 0.65  \\
K20  & Esophagitis    & 4515& 203501   & 0.66  \\
K22  & Other diseases of esophagus   & 6132& 201884   & 0.67  \\
K23  & Disorders of esophagus in diseases classified elsewhere& 4   & 208012   & 0.98  \\
K26  & Duodenal ulcer & 687 & 207329   & 0.70  \\
K28  & Gastrojejunal ulcer & 610 & 207406   & 0.71  \\
K40  & Inguinal hernia& 3585& 204431   & 0.70  \\
K44  & Diaphragmatic hernia& 2484& 205532   & 0.67  \\
K46  & Unspecified abdominal hernia  & 898 & 207118   & 0.67  \\
K50  & Crohn's disease  & 1898 & 206118   & 0.81  \\
K65  & Peritonitis    & 2788& 205228   & 0.66  \\
K67  & Disorders of peritoneum in infectious diseases classified elsewhere   & 13  & 208003   & 0.96  \\
K70  & Alcoholic liver disease  & 2726& 205290   & 0.74  \\
K71  & Toxic liver disease & 374 & 207642   & 0.68  \\
K72  & Hepatic failure, not elsewhere classified    & 1642& 206374   & 0.76  \\
K74  & Fibrosis and cirrhosis of liver    & 4888& 203128   & 0.71  \\
K83  & Other diseases of biliary tract    & 3251& 204765   & 0.66  \\
K85  & Acute pancreatitis  & 4293& 203723   & 0.66  \\
K90  & Intestinal malabsorption & 2428& 205588   & 0.83  \\
K91  & Intraoperative and postprocedural complications and disorders of digestive system, not elsewhere classified   & 1772& 206244   & 0.65  \\
K92  & Other diseases of digestive system & 18141    & 189875   & 0.66  \\
K94  & Complications of artificial openings of the digestive system& 2136& 205880   & 0.70  \\
K95  & Complications of bariatric procedures\\

\midrule
\rowcolor{gray!20} 
\multicolumn{5}{l}{XII. Diseases of the skin and subcutaneous tissue} \\
   
L11  & Other acantholytic disorders  & 400 & 207616   & 0.73  \\
L12  & Pemphigoid& 171 & 207845   & 0.90  \\
L22  & Diaper dermatitis   & 53  & 207963   & 0.67  \\
L27  & Dermatitis due to substances taken internally& 2264& 205752   & 0.68  \\
L42  & Pityriasis rosea    & 32  & 207984   & 0.68  \\
L43  & Lichen planus  & 573 & 207443   & 0.69  \\
L52  & Erythema nodosum    & 66  & 207950   & 0.69  \\
L57  & Skin changes due to chronic exposure to nonionizing radiation    & 16715    & 191301   & 0.70  \\
L59  & Other disorders of skin and subcutaneous tissue related to radiation  & 200 & 207816   & 0.67  \\
L60  & Nail disorders & 9636& 198380   & 0.66  \\
L62  & Nail disorders in diseases classified elsewhere   & 1   & 208015   & 0.86  \\
L65  & Other nonscarring hair loss   & 3892& 204124   & 0.66  \\
L68  & Hypertrichosis & 350 & 207666   & 0.65  \\
L70  & Acne & 2870& 205146   & 0.66  \\
L71  & Rosacea   & 3160& 204856   & 0.67  \\
L73  & Other follicular disorders    & 3652& 204364   & 0.72  \\
L82  & Seborrheic keratosis& 19887    & 188129   & 0.69  \\
L83  & Acanthosis nigricans& 134 & 207882   & 0.69  \\
L84  & Corns and callosities    & 4061& 203955   & 0.70  \\
L87  & Transepidermal elimination disorders    & 3   & 208013   & 0.66  \\
L89  & Pressure ulcer & 4029& 203987   & 0.69  \\
L94  & Other localized connective tissue disorders  & 862 & 207154   & 0.67  \\
L97  & Non-pressure chronic ulcer of lower limb, not elsewhere classified    & 6677& 201339   & 0.72  \\

\midrule
\rowcolor{gray!20} 
\multicolumn{5}{l}{XIII. Diseases of the musculoskeletal system and connective tissue} \\

M01  & Direct infections of joint in infectious and parasitic diseases classified elsewhere & 164 & 207852   & 0.92  \\
M07  & Enteropathic arthropathies    & 29  & 207987   & 0.77  \\
M08  & Juvenile arthritis  & 61  & 207955   & 0.68  \\
M10  & Gout & 8135& 199881   & 0.78  \\
M11  & Other crystal arthropathies   & 1152& 206864   & 0.68  \\
M12  & Other and unspecified arthropathy  & 13209    & 194807   & 0.65  \\
M14  & Arthropathies in other diseases classified elsewhere   & 448 & 207568   & 0.72  \\
M15  & Polyosteoarthritis  & 5637& 202379   & 0.69  \\
M16  & Osteoarthritis of hip    & 6221& 201795   & 0.66  \\
M17  & Osteoarthritis of knee   & 13448    & 194568   & 0.69  \\
M18  & Osteoarthritis of first carpometacarpal joint& 1806& 206210   & 0.67  \\
M19  & Other and unspecified osteoarthritis    & 17952    & 190064   & 0.66  \\
M1A  & Chronic gout   & 4508& 203508   & 0.74  \\
M22  & Disorder of patella & 848 & 207168   & 0.72  \\
M24  & Other specific joint derangements  & 1681& 206335   & 0.69  \\
M32  & Systemic lupus erythematosus (SLE) & 1676& 206340   & 0.65  \\
M33  & Dermatopolymyositis & 450 & 207566   & 0.78  \\
M34  & Systemic sclerosis {[}scleroderma{]}    & 627 & 207389   & 0.72  \\
M36  & Systemic disorders of connective tissue in diseases classified elsewhere   & 24  & 207992   & 0.75  \\
M40  & Kyphosis and lordosis    & 651 & 207365   & 0.70  \\
M48  & Other spondylopathies    & 12891    & 195125   & 0.66  \\
M61  & Calcification and ossification of muscle& 34  & 207982   & 0.71  \\
M66  & Spontaneous rupture of synovium and tendon   & 303 & 207713   & 0.67  \\
M80  & Osteoporosis with current pathological fracture   & 3189& 204827   & 0.69  \\
M81  & Osteoporosis without current pathological fracture& 12985    & 195031   & 0.70  \\
M83  & Adult osteomalacia  & 80  & 207936   & 0.80  \\
M85  & Other disorders of bone density and structure& 10859    & 197157   & 0.67  \\
M86  & Osteomyelitis  & 4033& 203983   & 0.67  \\
M88  & Osteitis deformans {[}Paget's disease of bone{]}  & 61  & 207955   & 0.77  \\
M89  & Other disorders of bone  & 14272    & 193744   & 0.79  \\
M97  & Periprosthetic fracture around internal prosthetic joint    & 389 & 207627   & 0.74  \\

\midrule
\rowcolor{gray!20} 
\multicolumn{5}{l}{XIV. Diseases of the genitourinary system} \\
N00  & Acute nephritic syndrome & 7   & 208009   & 0.91  \\
N01  & Rapidly progressive nephritic syndrome  & 41  & 207975   & 0.81  \\
N02  & Recurrent and persistent hematuria & 370 & 207646   & 0.80  \\
N03  & Chronic nephritic syndrome    & 116 & 207900   & 0.85  \\
N04  & Nephrotic syndrome  & 716 & 207300   & 0.71  \\
N05  & Unspecified nephritic syndrome& 1322& 206694   & 0.85  \\
N06  & Isolated proteinuria with specified morphological lesion    & 478 & 207538   & 0.68  \\
N07  & Hereditary nephropathy, not elsewhere classified  & 8   & 208008   & 0.93  \\
N08  & Glomerular disorders in diseases classified elsewhere  & 4897& 203119   & 0.75  \\
N10  & Acute pyelonephritis& 1131& 206885   & 0.70  \\
N11  & Chronic tubulo-interstitial nephritis   & 30  & 207986   & 0.70  \\
N12  & Tubulo-interstitial nephritis, not specified as acute or chronic & 3491& 204525   & 0.67  \\
N14  & Drug- and heavy-metal-induced tubulo-interstitial and tubular conditions   & 147 & 207869   & 0.90  \\
N15  & Other renal tubulo-interstitial diseases& 123 & 207893   & 0.70  \\
N17  & Acute kidney failure& 29637    & 178379   & 0.70  \\
N18  & Chronic kidney disease (CKD)  & 35840    & 172176   & 0.77  \\
N19  & Unspecified kidney failure    & 1978& 206038   & 0.82  \\
N21  & Calculus of lower urinary tract    & 598 & 207418   & 0.65  \\
N25  & Disorders resulting from impaired renal tubular function    & 3173& 204843   & 0.71  \\
N29  & Other disorders of kidney and ureter in diseases classified elsewhere & 37  & 207979   & 0.67  \\
N31  & Neuromuscular dysfunction of bladder, not elsewhere classified   & 3342& 204674   & 0.75  \\
N33  & Bladder disorders in diseases classified elsewhere& 5   & 208011   & 0.79  \\
N35  & Urethral stricture  & 430 & 207586   & 0.88  \\
N40  & Benign prostatic hyperplasia  & 14094    & 193922   & 0.85  \\
N41  & Inflammatory diseases of prostate  & 1115& 206901   & 0.69  \\
N42  & Other and unspecified disorders of prostate  & 436 & 207580   & 0.67  \\
N43  & Hydrocele and spermatocele    & 495 & 207521   & 0.71  \\
N45  & Orchitis and epididymitis& 655 & 207361   & 0.66  \\
N46  & Male infertility    & 235 & 207781   & 0.73  \\
N47  & Disorders of prepuce& 759 & 207257   & 0.71  \\
N48  & Other disorders of penis & 1496& 206520   & 0.68  \\
N49  & Inflammatory disorders of male genital organs, not elsewhere classified    & 341 & 207675   & 0.67  \\
N50  & Other and unspecified disorders of male genital organs & 2481& 205535   & 0.67  \\
N52  & Male erectile dysfunction& 6547& 201469   & 0.71  \\
N60  & Benign mammary dysplasia & 966 & 207050   & 0.71  \\
N61  & Inflammatory disorders of breast   & 529 & 207487   & 0.68  \\
N63  & Unspecified lump in breast    & 3650& 204366   & 0.65  \\
N64  & Other disorders of breast& 4596& 203420   & 0.67  \\
N65  & Deformity and disproportion of reconstructed breast    & 18  & 207998   & 0.87  \\
N70  & Salpingitis and oophoritis    & 177 & 207839   & 0.71  \\
N71  & Inflammatory disease of uterus, except cervix& 282 & 207734   & 0.82  \\
N73  & Other female pelvic inflammatory diseases    & 311 & 207705   & 0.74  \\
N75  & Diseases of Bartholin's gland & 92  & 207924   & 0.75  \\
N76  & Other inflammation of vagina and vulva  & 2934& 205082   & 0.73  \\
N77  & Vulvovaginal ulceration and inflammation in diseases classified elsewhere  & 6   & 208010   & 0.72  \\
N80  & Endometriosis  & 862 & 207154   & 0.82  \\
N81  & Female genital prolapse  & 1933& 206083   & 0.72  \\
N82  & Fistulae involving female genital tract & 361 & 207655   & 0.71  \\
N83  & Noninflammatory disorders of ovary, fallopian tube and broad ligament & 2586& 205430   & 0.74  \\
N84  & Polyp of female genital tract & 566 & 207450   & 0.71  \\
N85  & Other noninflammatory disorders of uterus, except cervix    & 674 & 207342   & 0.71  \\
N87  & Dysplasia of cervix uteri& 379 & 207637   & 0.74  \\
N88  & Other noninflammatory disorders of cervix uteri   & 450 & 207566   & 0.67  \\
N89  & Other noninflammatory disorders of vagina    & 5529& 202487   & 0.70  \\
N90  & Other noninflammatory disorders of vulva and perineum  & 1579& 206437   & 0.68  \\
N91  & Absent, scanty and rare menstruation    & 1476& 206540   & 0.80  \\
N92  & Excessive, frequent and irregular menstruation    & 2918& 205098   & 0.80  \\
N93  & Other abnormal uterine and vaginal bleeding  & 3578& 204438   & 0.73  \\
N94  & Pain and other conditions associated with female genital organs and menstrual cycle  & 3583& 204433   & 0.72  \\
N95  & Menopausal and other perimenopausal disorders& 7587& 200429   & 0.71  \\
N96  & Recurrent pregnancy loss & 149 & 207867   & 0.79  \\
N97  & Female infertility  & 808 & 207208   & 0.83  \\
N98  & Complications associated with artificial fertilization & 21  & 207995   & 0.93  \\
\midrule
\rowcolor{gray!20} 
\multicolumn{5}{l}{XV. Pregnancy, childbirth and the puerperium} \\
O00  & Ectopic pregnancy   & 55  & 207961   & 0.89  \\
O02  & Other abnormal products of conception   & 150 & 207866   & 0.84  \\
O03  & Spontaneous abortion& 241 & 207775   & 0.89  \\
O08  & Complications following ectopic and molar pregnancy    & 1   & 208015   & 0.84  \\
O09  & Supervision of high risk pregnancy & 2019& 205997   & 0.85  \\
O10  & Pre-existing hypertension complicating pregnancy, childbirth and the puerperium & 194 & 207822   & 0.67  \\
O11  & Pre-existing hypertension with pre-eclampsia & 21  & 207995   & 0.81  \\
O12  & Gestational {[}pregnancy-induced{]} edema and proteinuria without hypertension  & 61  & 207955   & 0.90  \\
O13  & Gestational {[}pregnancy-induced{]} hypertension without significant proteinuria& 297 & 207719   & 0.82  \\
O14  & Pre-eclampsia  & 353 & 207663   & 0.85  \\
O15  & Eclampsia & 104 & 207912   & 0.85  \\
O16  & Unspecified maternal hypertension  & 507 & 207509   & 0.80  \\
O20  & Hemorrhage in early pregnancy & 303 & 207713   & 0.84  \\
O21  & Excessive vomiting in pregnancy    & 395 & 207621   & 0.84  \\
O22  & Venous complications and hemorrhoids in pregnancy & 533 & 207483   & 0.66  \\
O23  & Infections of genitourinary tract in pregnancy    & 165 & 207851   & 0.82  \\
O24  & Diabetes mellitus in pregnancy, childbirth, and the puerperium   & 495 & 207521   & 0.83  \\
O25  & Malnutrition in pregnancy, childbirth and the puerperium    & 96  & 207920   & 0.90  \\
O26  & Maternal care for other conditions predominantly related to pregnancy & 766 & 207250   & 0.89  \\
O28  & Abnormal findings on antenatal screening of mother& 291 & 207725   & 0.74  \\
O30  & Multiple gestation  & 99  & 207917   & 0.89  \\
O32  & Maternal care for malpresentation of fetus   & 84  & 207932   & 0.91  \\
O34  & Maternal care for abnormality of pelvic organs    & 206 & 207810   & 0.87  \\
O35  & Maternal care for known or suspected fetal abnormality and damage& 191 & 207825   & 0.78  \\
O36  & Maternal care for other fetal problems  & 965 & 207051   & 0.86  \\
O40  & Polyhydramnios & 48  & 207968   & 0.92  \\
O41  & Other disorders of amniotic fluid and membranes   & 172 & 207844   & 0.78  \\
O42  & Premature rupture of membranes& 52  & 207964   & 0.82  \\
O43  & Placental disorders & 29  & 207987   & 0.90  \\
O44  & Placenta previa& 141 & 207875   & 0.88  \\
O45  & Premature separation of placenta {[}abruptio placentae{]}   & 15  & 208001   & 0.81  \\
O46  & Antepartum hemorrhage, not elsewhere classified   & 221 & 207795   & 0.82  \\
O47  & False labor    & 80  & 207936   & 0.79  \\
O48  & Late pregnancy & 51  & 207965   & 0.88  \\
O60  & Preterm labor  & 63  & 207953   & 0.91  \\
O69  & Labor and delivery complicated by umbilical cord complications   & 8   & 208008   & 0.82  \\
O70  & Perineal laceration during delivery& 13  & 208003   & 0.93  \\
O71  & Other obstetric trauma   & 7   & 208009   & 0.89  \\
O72  & Postpartum hemorrhage    & 47  & 207969   & 0.87  \\
O73  & Retained placenta and membranes, without hemorrhage    & 20  & 207996   & 0.89  \\
O76  & Abnormality in fetal heart rate and rhythm complicating labor and delivery & 33  & 207983   & 0.71  \\
O80  & Encounter for full-term uncomplicated delivery    & 692 & 207324   & 0.88  \\
O82  & Encounter for cesarean delivery without indication& 611 & 207405   & 0.83  \\
O85  & Puerperal sepsis    & 94  & 207922   & 0.67  \\
O86  & Other puerperal infections    & 131 & 207885   & 0.68  \\
O87  & Venous complications and hemorrhoids in the puerperium & 5   & 208011   & 0.79  \\
O88  & Obstetric embolism  & 35  & 207981   & 0.82  \\
O89  & Complications of anesthesia during the puerperium & 1   & 208015   & 0.92  \\
O90  & Complications of the puerperium, not elsewhere classified   & 264 & 207752   & 0.72  \\
O91  & Infections of breast associated with pregnancy, the puerperium and lactation    & 20  & 207996   & 0.84  \\
O92  & Other disorders of breast and disorders of lactation associated with pregnancy and the puerperium   & 62  & 207954   & 0.88  \\
O98  & Maternal infectious and parasitic diseases classifiable elsewhere but complicating pregnancy, childbirth and the puerperium  & 235 & 207781   & 0.86  \\
O99  & Other maternal diseases classifiable elsewhere but complicating pregnancy, childbirth and the puerperium & 1245& 206771   & 0.85  \\
O9A  & Maternal malignant neoplasms, traumatic injuries and abuse classifiable elsewhere but complicating pregnancy, childbirth and the puerperium & 10  & 208006   & 0.93  \\
P00  & Newborn affected by maternal conditions that may be unrelated to present pregnancy   & 5   & 208011   & 0.97  \\
P01  & Newborn affected by maternal complications of pregnancy& 3   & 208013   & 0.93  \\
P02  & Newborn affected by complications of placenta, cord and membranes& 3   & 208013   & 0.85  \\
P03  & Newborn affected by other complications of labor and delivery    & 2   & 208014   & 0.75  \\
P04  & Newborn affected by noxious substances transmitted via placenta or breast milk  & 42  & 207974   & 0.69  \\
P05  & Disorders of newborn related to slow fetal growth and fetal malnutrition   & 56  & 207960   & 0.66  \\
P07  & Disorders of newborn related to short gestation and low birth weight, not elsewhere classified & 55  & 207961   & 0.91  \\
P08  & Disorders of newborn related to long gestation and high birth weight  & 11  & 208005   & 0.87  \\
P09  & Abnormal findings on neonatal screening & 43  & 207973   & 0.88  \\
P12  & Birth injury to scalp    & 28  & 207988   & 0.72  \\
P22  & Respiratory distress of newborn    & 3   & 208013   & 0.99  \\
P24  & Neonatal aspiration & 72  & 207944   & 0.73  \\
P27  & Chronic respiratory disease originating in the perinatal period  & 33  & 207983   & 0.87  \\
P28  & Other respiratory conditions originating in the perinatal period & 10  & 208006   & 0.89  \\
P29  & Cardiovascular disorders originating in the perinatal period& 113 & 207903   & 0.75  \\
P36  & Bacterial sepsis of newborn   & 1   & 208015   & 1.00  \\
P37  & Other congenital infectious and parasitic diseases& 2   & 208014   & 0.87  \\
P38  & Omphalitis of newborn    & 50  & 207966   & 0.90  \\
P52  & Intracranial nontraumatic hemorrhage of newborn   & 1   & 208015   & 0.95  \\
P59  & Neonatal jaundice from other and unspecified causes    & 10  & 208006   & 0.99  \\
P61  & Other perinatal hematological disorders & 4   & 208012   & 1.00  \\
P70  & Transitory disorders of carbohydrate metabolism specific to newborn   & 3   & 208013   & 1.00  \\
P74  & Other transitory neonatal electrolyte and metabolic disturbances & 20  & 207996   & 0.69  \\
P77  & Necrotizing enterocolitis of newborn    & 1   & 208015   & 1.00  \\
P78  & Other perinatal digestive system disorders   & 11  & 208005   & 0.99  \\
P81  & Other disturbances of temperature regulation of newborn& 4   & 208012   & 0.99  \\
P83  & Other conditions of integument specific to newborn& 47  & 207969   & 0.87  \\
P90  & Convulsions of newborn   & 12  & 208004   & 0.96  \\
P91  & Other disturbances of cerebral status of newborn  & 18  & 207998   & 0.90  \\
P92  & Feeding problems of newborn   & 53  & 207963   & 0.99  \\
P94  & Disorders of muscle tone of newborn& 4   & 208012   & 0.97  \\
P95  & Stillbirth& 2   & 208014   & 0.71  \\
P96  & Other conditions originating in the perinatal period   & 34  & 207982   & 0.90 
\end{longtable}

\end{document}